\documentclass[useAMS,usenatbib]{mnras}
\usepackage{pdflscape}
\usepackage{graphicx}
\usepackage{txfonts}

\newcommand{\mytilde}{\raise.17ex\hbox{$\scriptstyle\mathtt{\sim}$}}

\usepackage[T1]{fontenc}
\usepackage{ae,aecompl}

\title[Kepler's Habitable Host Stars]{The Host Stars of \emph{Kepler}'s Habitable Exoplanets: Superflares, Rotation and Activity}

\author[Armstrong et. al.]{\parbox{\textwidth}{D. J. Armstrong$^{1,2}$\thanks{d.j.armstrong@warwick.ac.uk}, C. E. Pugh$^1$, A.-M. Broomhall$^{1}$, D. J. A. Brown$^1$, M. N. Lund$^{3}$, H. P. Osborn$^1$, D. L. Pollacco$^1$}
\vspace{0.4cm}\\
\parbox{\textwidth}{$^{1}$University of Warwick, Department of Physics, Gibbet Hill Road, Coventry, CV4 7AL, UK\\
$^{2}$ARC, School of Mathematics \& Physics, Queen's University Belfast, University Road, Belfast BT7 1NN, UK\\
$^{3}$Stellar Astrophysics Centre (SAC), Department of Physics and Astronomy, Aarhus University, Ny Munkegade 120, DK-8000 Aarhus C, Denmark}}

\date{Accepted . Received}

\pubyear{2015}

\begin{document}

\pagerange{\pageref{firstpage}--\pageref{lastpage}}

\maketitle

\begin{abstract}
We embark on a detailed study of the lightcurves of \emph{Kepler}'s most Earth-like exoplanet host stars using the full length of \emph{Kepler} data. We derive rotation periods, photometric activity indices, flaring energies, mass loss rates, gyrochronological ages, X-ray luminosities and consider implications for the planetary magnetospheres and habitability. Furthermore, we present the detection of superflares in the lightcurve of Kepler-438, the exoplanet with the highest Earth Similarity Index to date. Kepler-438b orbits at a distance of 0.166AU to its host star, and hence may be susceptible to atmospheric stripping. Our sample is taken from the Habitable Exoplanet Catalogue, and consists of the stars Kepler-22, Kepler-61, Kepler-62, Kepler-174, Kepler-186, Kepler-283, Kepler-296, Kepler-298, Kepler-438, Kepler-440, Kepler-442, Kepler-443 and KOI-4427, between them hosting 15 of the most habitable transiting planets known to date from \emph{Kepler}.
\end{abstract}


\section{Introduction}

In recent years the pace of discovery of exoplanets has intensified, with an increasing number of small, potentially rocky planets being found. This is largely due to the \emph{Kepler} mission \citep{Borucki:2010dn,2010ApJ...713L..79K}, which studies \mytilde 150000 stars with high-precision photometry, and has found several thousand candidate exoplanets. With more planets comes a focus on new questions, including the potential habitability of these planets in a much more diverse range of environments than known in the Solar System. 

The key driver of an exoplanet's local environment is its host star. This leads to well-known properties such as the equilibrium temperature of the exoplanet, defining the habitable zone where liquid water could exist on the planet's surface \citep[e.g.][]{Kasting:1993hw}. Many of the newly discovered planets orbit stars cooler than the Sun, because around these stars habitable zone planets are often easier to detect. These stars are known to have increased activity relative to Sun-like stars \citep{Wright:2011dj}, as well as increased potential for flaring, coronal mass ejections (CMEs), X-ray and EUV flux \citep{Mamajek:2008jz}, and active magnetic fields \citep{Reiners:2012ce,Vidotto:tt}. Such properties tend to get weaker as a host star ages and spins down \citep[e.g.][along with many others]{Hempelmann:1995uy,vanSaders:2013bs,Garcia:2014ds} and all have a potential effect on planetary habitability.

Stellar flares are associated with increased UV and charged particle flux, but this is thought not to affect planetary habitability \citep{2010AsBio..10..751S}. They are, however, also associated with increased likelihood of CMEs \citep{Chen:2010gz}, which can compress planetary magnetospheres \citep{Khodachenko:2007hm} and drive atmospheric erosion \citep{Lammer:2007kh}. UV flux, whether from a flare or background stellar radiation, can affect atmospheric heating and chemistry, as well as changing the biomarkers which future missions might search for \citep{Grenfell:2014gy,France:2014bh}. Stellar activity is associated with increased flaring rates, also potentially impacting atmospheric biomarkers and in some strong cases destroying ozone, an important element in shielding the Earth from radiation \citep{JohnLeeGrenfell:2012fa}.

Planetary magnetospheres are important for shielding planets from potential atmospheric erosion and from high-energy particles, with unshielded planets orbiting M dwarfs losing their atmospheres in 1 Gyr in some cases \citep{Zendejas:2010ib}. The size of a magnetosphere of a planet is strongly affected by the host star, in particular its stellar wind \citep{See:2014gk} and magnetic fields \citep{Vidotto:2013hz}. Variations in the stellar wind can interact strongly with a planet's atmosphere, stripping it or depositing heat and gravity waves \citep{Cohen:2014eb,Cohen:2015gd}. Considering the local galactic environment rather than the star could also have an effect; recently \citet{Johnson:2015up} showed that variable interstellar medium density can affect the astrosphere of planetary systems, changing the level of shielding to cosmic rays.

Exoplanets are often rated on the Earth Similarity Index \citep[ESI][]{SchulzeMakuch:2011cp}, which measures the similarity of a planet to Earth based on its radius, bulk density, escape velocity and surface temperature. In this work we consider the host stars of the most Earth-like exoplanets defined by this index, as found in the Habitable Exoplanets Catalogue\footnote{http://phl.upr.edu/projects/habitable-exoplanets-catalog}. To allow an in-depth photometric study of this sample, we limit ourselves to stars observed by the \emph{Kepler} mission, giving a sample of 13 stars hosting 15 highly ranked planets. The sample consisted of Kepler-22b  \citep[KIC 10593626;][]{Borucki:2012hw}, Kepler-61b \citep[KIC 6960913;][]{Ballard:2013ir}, Kepler-62e and f \citep[KIC 9002278;][]{Borucki:2013js}, Kepler-174d \citep[KIC 8017703;][]{Rowe:2014wz}, Kepler-186f  \citep[KIC 8120608;][]{Quintana:2014cc}, Kepler-283c (KIC 10604335), Kepler-296e and f (KIC 11497958), Kepler-298d \citep[KIC 11176127;][]{Rowe:2014wz}, Kepler-438b (KIC 6497146), Kepler-440b (KIC 6106282), Kepler-442b (KIC 4138008), Kepler-443b (KIC 11757451) and KOI-4427b \citep[KIC 4172805;][]{2015ApJ...800...99T}, where KOI stands for \emph{Kepler} Object of Interest, and KIC represents the Kepler Input Catalogue identifier for the star. Kepler-296 is a binary system with 5 transiting planets. It has only recently been shown that all these planets likely orbit the same star in the binary \citep{Barclay:2015wa}. The parameters of these stars are listed in Table \ref{tabinputparams}, in some cases from a more recent source than the discovery paper.

\section{Data}
All \emph{Kepler} lightcurves are available publicly via the Michulski Archive for Space Telescopes (MAST). These lightcurves are provided both in raw format and as the detrended PDC ms-MAP lightcurves \citep[Presearch Data Conditioning multi-scale Maximum A Posteriori;][]{Stumpe:2012bj,Smith:2012ji}, and span a range of approximately 4 years from May 2009 to May 2013. We make use of the full time series, comprising Quarters 0-17 (the spacecraft reorientated itself every \mytilde 90 days, separating the data into Quarters). The PDC lightcurves are explicitly designed to remove instrumental variability while preserving astrophysical flux variations. However, they also apply a high-pass filter to the data, effectively attenuating any signal on a timescale longer than approximately 21 days \citep{Garcia:2013td}. For work involving exoplanetary transits this usually has little effect, but for the study of long stellar rotation periods problems can be introduced. As a large part of this work derives from determinations of the host star rotation periods, we make use of the KASOC lightcurves. See \citet{Handberg:2014ih} for a detailed description of these lightcurves and how they are generated. These can be processed with filters of different durations, and in this work we employ lightcurves using filters of 30, 50, and 100 days. A shorter filter tends to produce cleaner lightcurves, but also attenuates signal on timescales longer than the filter. Before any further processing, transits of the known planets are cut using their published ephemeris. 

Each of the systems studied here have been investigated in depth in their respective discovery papers. Here we concentrate on the host stars, as well as the effects these may have on their orbiting planets.

\begin{table*}
\begin{center}
\caption{Host Star Input Parameters}
\label{tabinputparams}
\begin{tabular}{lrrrrrrl}
\hline
Star & T$_\textrm{eff}$ & Log g & Fe/H & R$_*$ & M$_*$ & $\rho_*$  & Source \\
& K & dex & dex & $R_\odot$ & $M_\odot$ & $gcm^{-3}$& \\
\hline
Kepler-22 & $5518 \pm 44$ & $4.44 \pm 0.06$ & $-0.29 \pm 0.06$ & $0.979 \pm 0.02$ & $0.970\pm0.06$ & $1.46\pm0.13^{a}$ & \citet{Borucki:2012hw} \\[2pt]
Kepler-61 & $4016^{+68}_{-150}$ & $4.66^{+0.08}_{-0.04}$ & $0.03 \pm 0.14$ & $0.62^{+0.02}_{-0.05}$ & $0.635\pm0.037$ & $3.76^{+0.94}_{-0.42}$$^{a}$ & \citet{Ballard:2013ir}\\[2pt]
Kepler-62 & $4925 \pm 70$ & $4.68 \pm 0.04$ & $-0.37 \pm 0.04$ & $0.64 \pm 0.02$ & $0.69\pm0.02$ & $3.72\pm0.36$$^{a}$ & \citet{Borucki:2013js}\\[2pt]
Kepler-174 & $4880 \pm 126$ & $4.679 \pm 0.15$ & $-0.43 \pm 0.10$ & $0.622 \pm 0.032$ & $0.67\pm0.13$$^{a}$ & $3.956 \pm0.450$& \citet{Rowe:2014wz}\\[2pt]
Kepler-186 & $3755 \pm 90$ & $4.736 \pm 0.02$ & $-0.26 \pm 0.12$ & $0.523^{+0.023}_{-0.021}$ & $0.544^{+0.024}_{-0.021}$ & $5.29^{+0.54}_{-0.39}$& \citet{2015ApJ...800...99T}\\[2pt]
Kepler-283 & $4351 \pm 100$ & $4.72 \pm 0.15$ & $-0.2 \pm 0.1$ & $0.566 \pm 0.024$ & $0.61\pm0.09$$^{a}$ &$4.770\pm0.403$ & \citet{Rowe:2014wz} \\[2pt]
Kepler-296 & $3740 \pm 130$ & $4.774^{+0.091}_{-0.059} $ & $-0.08 \pm 0.3$ & $0.48^{+0.066}_{-0.087}$ & $0.498^{+0.067}_{-0.087}$ & $6.4^{+3.2}_{-1.5}$& \citet{Barclay:2015wa} \\[2pt]
Kepler-298 & $4465 \pm 100$ & $4.709 \pm 0.15$ & $-0.24 \pm 0.1$ & $0.582 \pm 0.025$ & $0.63\pm0.10$$^{a}$ & $4.526\pm0.414$& \citet{Rowe:2014wz} \\[2pt]
Kepler-438 & $3748 \pm 112$ & $4.74^{+0.059}_{-0.029}$ & $0.16 \pm 0.14$ & $0.520^{+0.038}_{-0.061}$ & $0.544^{+0.041}_{-0.061}$ & $5.52^{+1.53}_{-0.77}$& \citet{2015ApJ...800...99T} \\[2pt]
Kepler-440 & $4134 \pm 154$ & $4.706^{+0.049}_{-0.016}$ & $-0.30 \pm 0.15$ & $0.559^{+0.029}_{-0.054}$ & $0.575^{+0.043}_{-0.047}$ &$4.76^{+1.03}_{-0.48}$ &  \citet{2015ApJ...800...99T}\\[2pt]
Kepler-442 & $4402 \pm 100$ & $4.673^{+0.018}_{-0.021}$ & $-0.37 \pm 0.10$ & $0.598^{+0.023}_{-0.024}$ & $0.609^{+0.03}_{-0.026}$ &$4.01^{+0.37}_{-0.30}$ & \citet{2015ApJ...800...99T} \\[2pt]
Kepler-443 & $4723 \pm 100$ & $4.614^{+0.016}_{-0.029}$ & $-0.01 \pm 0.10$ & $0.706^{+0.028}_{-0.024}$ & $0.738^{+0.033}_{-0.029}$ &$2.96^{+0.24}_{-0.25}$ & \citet{2015ApJ...800...99T} \\[2pt]
KOI-4427 & $3813 \pm 112$ & $4.751^{+0.067}_{-0.030}$ & $-0.07 \pm 0.14$ & $0.505^{+0.038}_{-0.065}$ & $0.526^{+0.040}_{-0.062}$ &$5.79^{+1.87}_{-0.82}$ & \citet{2015ApJ...800...99T} \\[2pt]
\hline
\multicolumn{8}{l}{$^a$ Derived from parameters in source paper}
\end{tabular}
\end{center}
\end{table*}

\section{Methods}
\subsection{Rotation Periods}
\begin{figure}
\resizebox{\hsize}{!}{\includegraphics{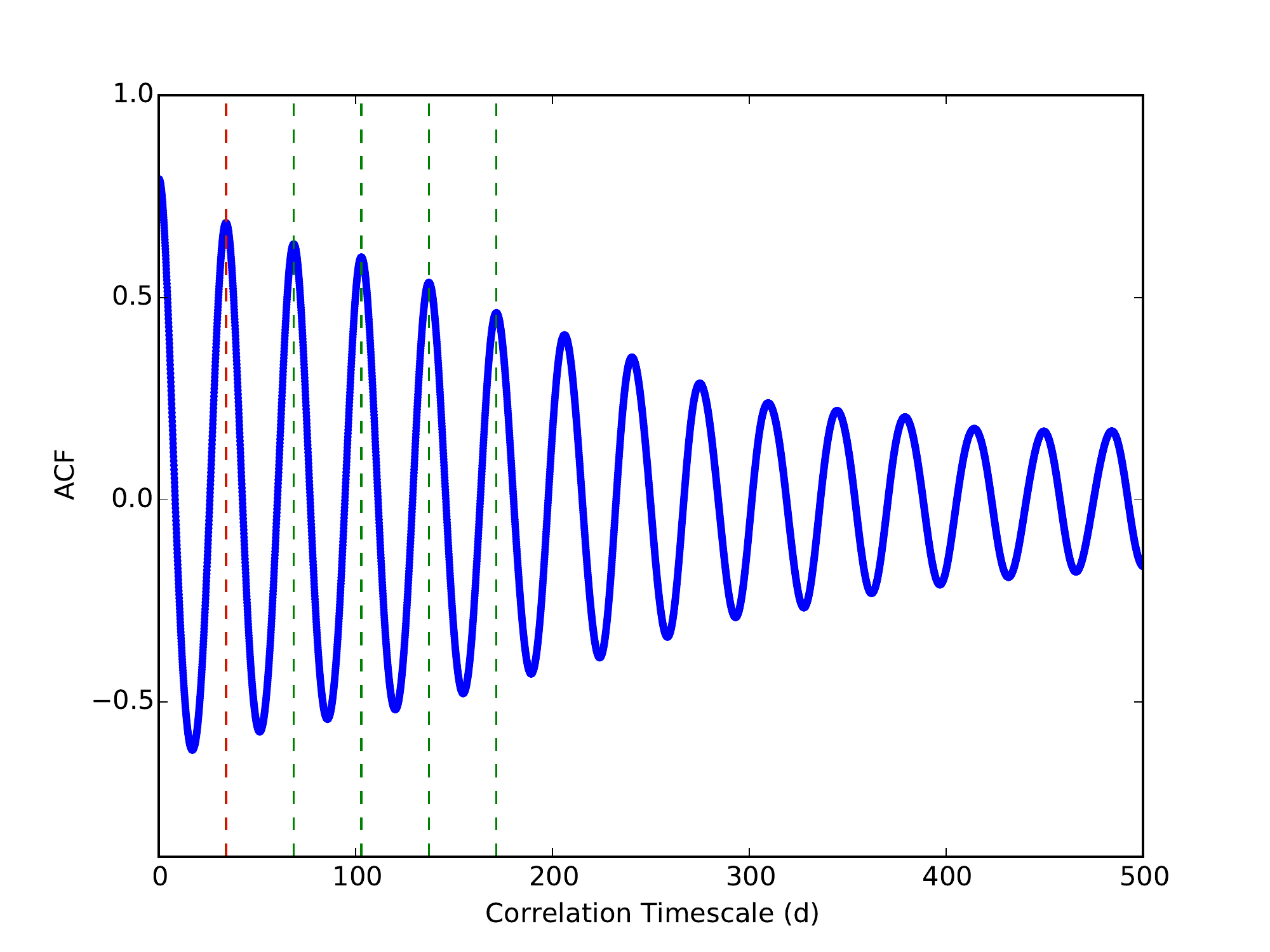}}
\caption{Auto correlation function for Kepler-186, calculated using the KASOC 50d lightcurve. The rotation period is marked as the leftmost dashed line, with the peaks used to fit and extract this period also marked. The ACF has been smoothed by a gaussian filter.}
\label{figkep186acf}
\end{figure}

Several of the target stars already have determinations of the rotation period. These were generally performed using PDC data, which as described above can lead to problems with long rotation periods. As such we independently determine stellar rotation periods using the KASOC data. We first consider the auto-correlation-function (ACF) of the lightcurves, a method which has been shown to robustly retrieve periodic signals in stellar lightcurves \citep{McQuillan:2013df}. We follow a similar procedure to \citet{McQuillan:2014gp} in calculating ACF periods. Initially gaps in the data are filled with zeros, which does not affect the ACF calculation. Data are then binned down by a factor of 4 cadences to speed processing. The ACF is then calculated via the equation:  
\begin{equation}
r_k = \frac{\sum\limits_{i=1}^{N-k}(x_i - \bar{x})(x_{i+k} - \bar{x})}{\sum\limits_{i=1}^{N}(x_i - \bar{x})^2}
\end{equation}
where the time series is given by $x_i (i = 1,...,N)$, $\bar{x}$ is its mean, and the ACF value $r_k$ is computed at successive lags $k$, with $k$ being an integer multiple of the cadence. Extracting a period from this curve is non-trivial; we first smooth the curve using a Gaussian filter with a standard deviation the same as the maximum peak (to avoid adversely affecting this peak). The filter is truncated at $3.1\sigma$. This leads to a curve of the form of Fig. \ref{figkep186acf}. We then identify the first 4 harmonics of the first peak in the curve, and perform a linear fit to the locations of these peaks, as well as the first peak and the origin, giving 6 points overall. The final gradient and error of this fit then gives our ACF period and its error. Due to our small sample number we are able to confirm the extracted period against the lightcurves themselves.

\begin{figure*}
\resizebox{\hsize}{!}{\includegraphics{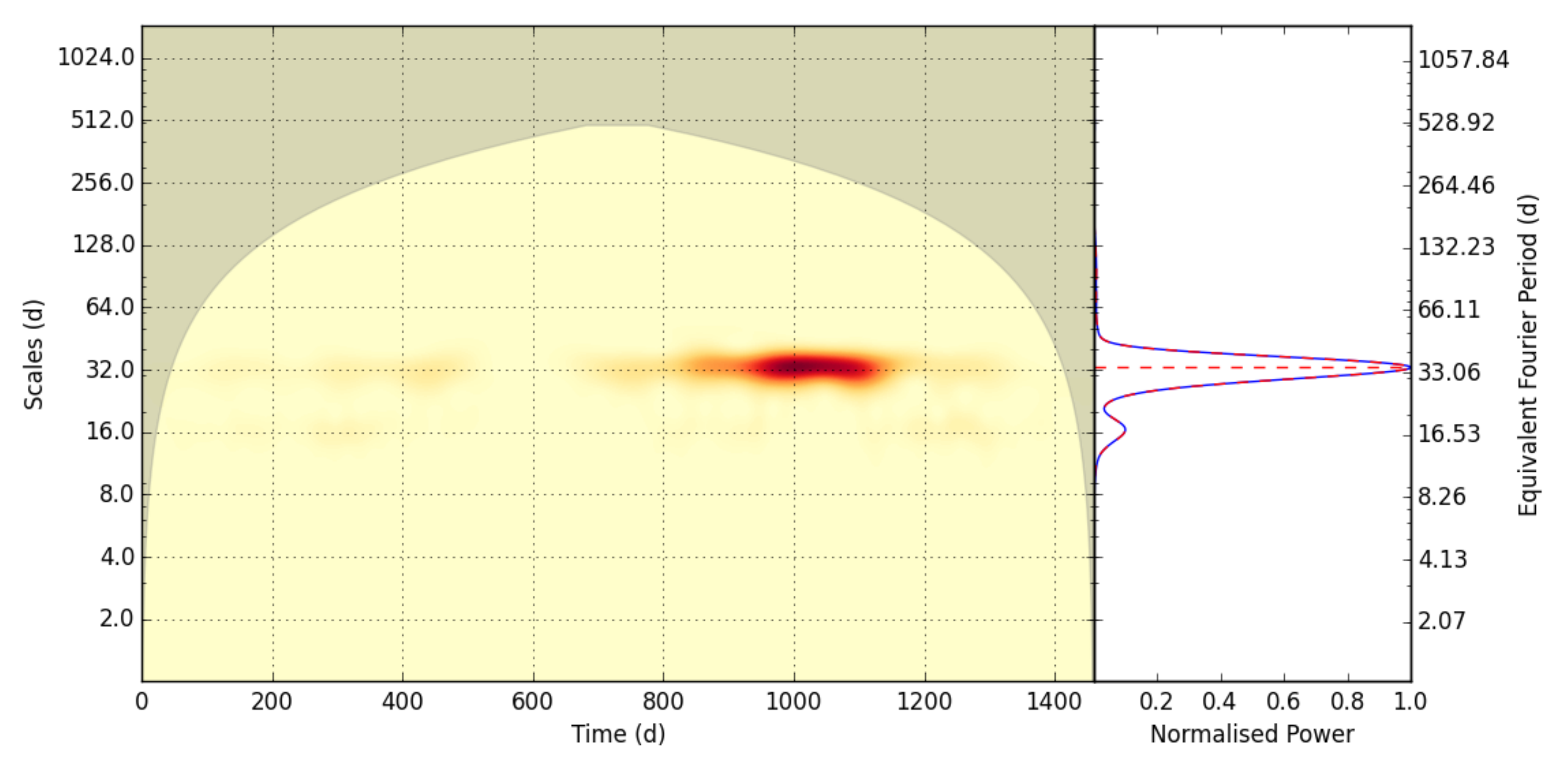}}
\caption{Contour wavelet plot for Kepler-186, calculated using the KASOC 50d lightcurve. The global wavelet spectrum is plotted to the right and fit by a sum of Gaussians (curved dashed line). The horizontal dashed line shows the location of the maximum peak seen in the GWS. The rotation period is visible throughout the data, but becomes especially strong during an extended active region near 1000d from the data start. The P/2 harmonic can also be clearly seen.}
\label{figkep186wav}
\end{figure*}

It is also useful to consider where this signal arises, as in some cases one noisy region of data can produce a false rotation signal. We therefore also perform a wavelet analysis \citep{Torrence:1998wk}, as has been used for \emph{Kepler} data before \citep[e.g.][]{Mathur:2014cz,Garcia:2014ds}. This allows a study of the rotation signal in the time domain, although giving reduced resolution in frequency. Wavelet analysis consists of convolving a selected waveform with the data at each timestep, with the convolution performed repeatedly using a range of scales for the waveform. Each scale represents a different frequency. The power given by each convolution then produces a map of frequency against time. We use a custom-built code, centred around the open source python module `wavelets' \footnote{https://github.com/aaren/wavelets}. We use a Morlet wavelet, as has been previously used for stellar rotation applications. Rather than filling data gaps with zeros, which could bias the calculation, we use linear interpolation between the average of the 20 points on either side of each gap. For gaps greater than 10 days, we continue to use zeros rather than assume a linear relation over an extended timespan. We use a scale resolution of 0.01, and unbias the resulting power spectrum as described in \citet{Liu:uw}. An example wavelet plot is shown in Fig. \ref{figkep186wav}.

To extract a period from the array of wavelet powers, the power is summed over the time axis, resulting in the global wavelet spectrum (GWS). The shape of a single period in the GWS is Gaussian, hence we approximate that the total GWS can be fit by the sum of several Gaussians \citep[following the method of][]{Garcia:2014ds}. The fit is initialised with a Gaussian assigned to each peak in the GWS. Each Gaussian is given an initial amplitude of the GWS value at that peak location, with a standard deviation of a quarter of the peak's period. The fit is then found via least-squares optimization. The extracted period is then the peak location of the highest amplitude Gaussian, and the error the Half-Width-Half-Maximum of this peak. We note that the GWS has degraded resolution compared to a Fourier transform. It tends to give larger uncertainties, as can be seen in our results and as has been seen previously \citep[e.g.][]{Garcia:2014ds}. This is partially because the GWS includes effects such as differential rotation, which can blur a periodic signal. However, following this same procedure for a test \emph{Kepler}-length lightcurve consisting of a pure 30 day sinusoid without noise still resulted in a period error of \mytilde 3 days, showing that the GWS produces an intrinsically large error even for clear periodicities. Hence in adopting final rotation values we use the ACF periods, but backed up by the wavelet spectrum. To ensure we do not pick up lower harmonics of the true rotation period due to attenuation, we adopt the period from the KASOC lightcurve with the shortest filter which is consistent with the 100 day KASOC lightcurve value (except for Kepler-62, see below).

The resulting stellar rotation periods are given in Table \ref{tabderivedparams}, and discussed in Section \ref{discprot}. The remaining wavelet plots are given in the Appendix. We are able to obtain good rotation periods for 6 stars. In addition Kepler-62 shows the rotation period given in \citet{Borucki:2013js} of near 39 days, but with a weak signal evident in only a few parts of the lightcurve, with only partial evidence backing this period up via visual inspection. We proceed using this period, but caution that it is unclear if this is the true rotation period. In addition the KASOC 100 day lightcurve for Kepler-62 shows a $111\pm9$ day period, too long to be clearly confirmed given the \emph{Kepler} 90 day quarter length but intriguing nevertheless. Kepler-174, Kepler-298, Kepler-442 and Kepler-443 give periods which are multiples of the 90 day quarter length and arise from artefacts associated with this. KOI-4427 gives a very weakly determined signal near 40 or 80 days. We are unable to distinguish between the two, and the signal arises from few regions of the lightcurve. Hence we do not consider this as a conclusive period determination.

\subsection{Stellar Photometric Activity}
We next study the photometric activity of these stars. A number of methods for investigating activity with the \emph{Kepler} lightcurves have been proposed, including the Range \citep{Basri:2010kk,Basri:2013fi}, S$_{phot}$ \citep{Garcia:2010gh} and more recently S$_{phot,k}$ \citep{Mathur:2014cz,Garcia:2014ds}. The latter incorporates the stellar rotation period into the measurement of activity, an important link given that activity usually appears on timescales of the stellar rotation period due to spot modulation. It has also been shown that the range can underestimate the true variability \citep{Garcia:2014ds}. As such we utilise S$_{phot,k}$ for this study. The choice of lightcurve used to calculate the activity is important. We trialled the PDC and each KASOC lightcurve, and found that as expected the measured activity is least for the PDC lightcurves due to the stronger attenuation of these generally longer period spot modulation signals. This trend continued through the KASOC lightcurves, with the highest activity measures found for the 100 day KASOC lightcurves. Finding the point at which this increased activity becomes due to instrumental effects which have not been removed is non-trivial. Given the \mytilde 30 day rotation periods studied here, we chose to use the 50 day KASOC lightcurves for studying activity. We used these even for the 2 stars with shorter than 30 day rotation periods, for consistency. These lightcurves should preserve signals at the rotation periods of the stars well, without introducing the additional chance for instrumental noise of the 100 day KASOC lightcurves.

\begin{figure}
\resizebox{\hsize}{!}{\includegraphics{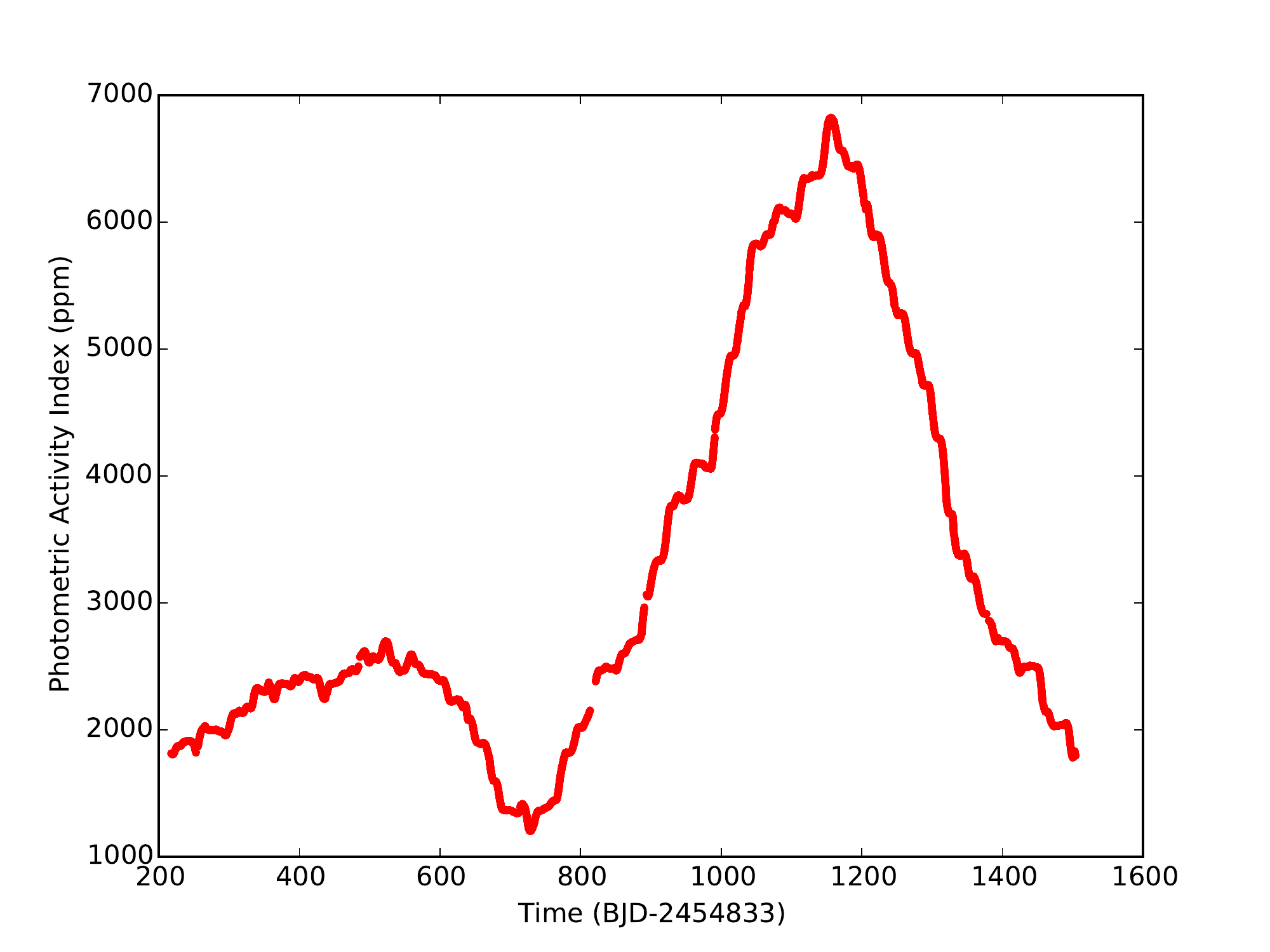}}
\caption{Photometric Activity Index $S_{phot,5}$ for Kepler-186, calculated using the ACF rotation period of 34.27 days. The active region evident in Fig. \ref{figkep186wav} is clearly visible.}
\label{figkep186sphot}
\end{figure}

To calculate S$_{phot,k}$, the lightcurve is divided into successive windows of length $k$ multiplied by the stellar rotation period. It has been shown that a robust  indicator for magnetic activity is given when using $k=5$ \citep{Mathur:2014cz}, which we adopt here. The standard deviation of each window is taken, and the final measure $\left<S_{phot,5}\right>$ given by the mean of these. Windows containing less than 3/4 of the expected number of data points in a 5 rotation period timespan (due to, for example, gaps in the data) were not included. Where we could obtain no good rotation period for a star we performed this analysis using a period of 30 days, to allow some measure of the activity.   Derived values for $\left<S_{phot,5}\right>$ are given in Table \ref{tabderivedparams}, and discussed in Section \ref{discactivity}. By plotting the value of $S_{phot,5}$ for each window it is possible to track the activity of the star through the 4 years of \emph{Kepler} data, an example of which is given in Fig. \ref{figkep186sphot}. We do not see any strong evidence for activity cycles in any of our targets. We also calculate the Contrast for each star, as defined in \citep{Mathur:2014cz}. The Contrast is the ratio of the average value of $S_{phot,5}$ in the windows with greater than the global value of $S_{phot}$ (i.e. the standard deviation of the whole dataset) to that of the windows with lower than the global value, and gives a measure of the activity variation seen in the data.

\subsection{Flares}

We inspected each of the sample lightcurves (in long and short cadence) for evidence of flares, which have been detected in \emph{Kepler} data previously for other stars \citep{2012Natur.485..478M}. For this study we utilised the PDC lightcurves in both long and short cadence, as flares occur on a timescale of \mytilde a few cadences in the long cadence data, easily avoiding attenuation by the PDC detrending. For one star, Kepler-438, 7 significant flares were found during the 4 years of \emph{Kepler} data. Unfortunately no short cadence data is available for Kepler-438, but the flares are easily discernible in the long cadence data. Estimates of the flare energies were made using the flare amplitude, duration, and luminosity, and were done in a similar manner to \citet{2013ApJS..209....5S}. 

\begin{figure}
\resizebox{\hsize}{!}{\includegraphics{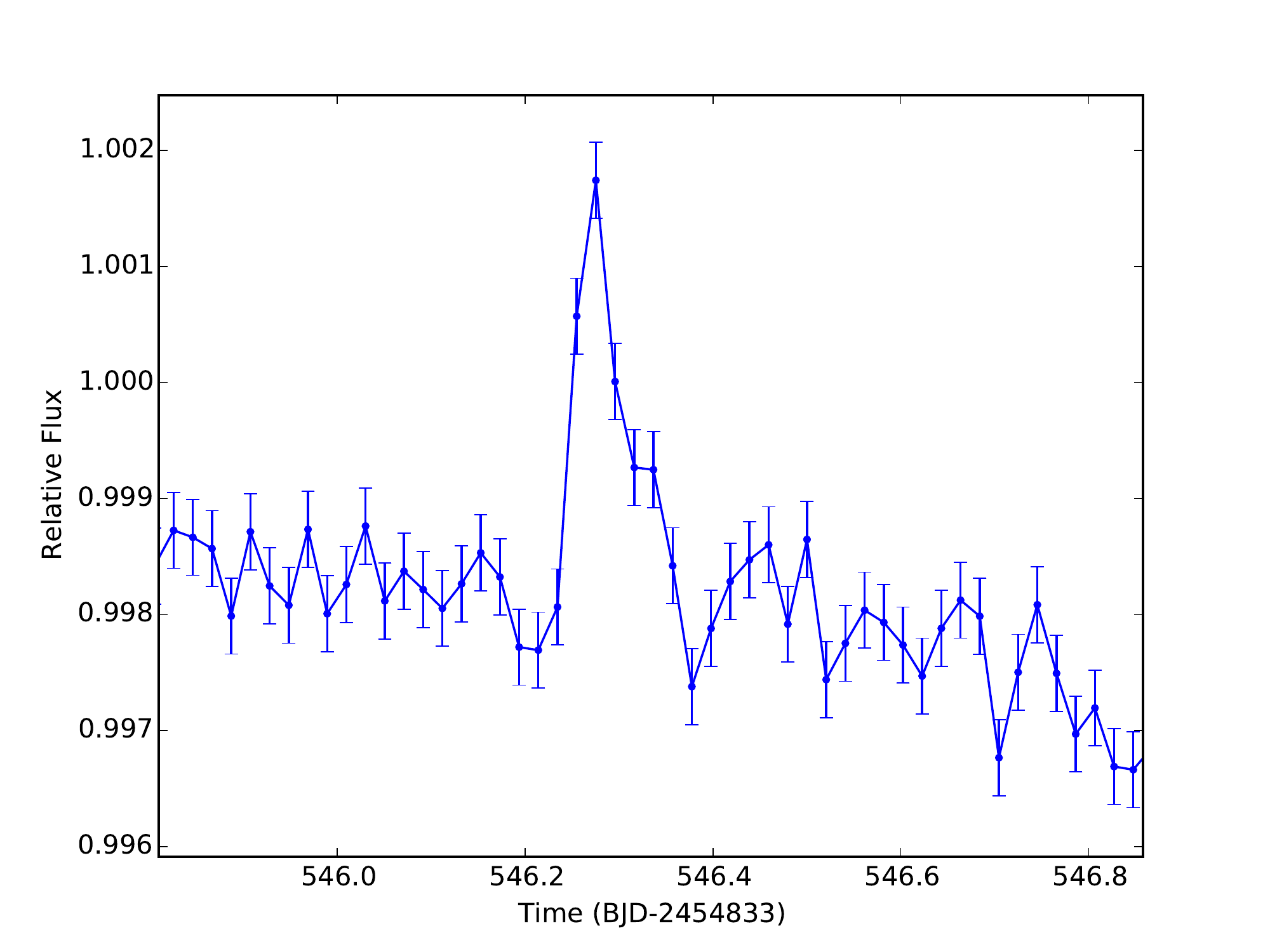}}
\caption{Superflare seen in Quarter 6 of the Kepler-438 lightcurve, with an energy of \mytilde $10^{33}$ erg}
\label{figkep438flare}
\end{figure}

The total energy of the flare can be estimated by integrating the bolometric luminosity of the flare over the flare duration:
\begin{equation}
	E_{flare} = \int_{flare} L_{flare}(t)\,\mathrm{d}t.
\end{equation}
Assuming that the flare emission can be approximated by a blackbody spectrum, the luminosity of the flare is given by:
\begin{equation}
	L_{flare} = \sigma_{SB} T^{4}_{flare}A_{flare},
\end{equation}
where $\sigma_{SB}$ is the Stefan-Boltzmann constant, $T_{flare}$ is the effective temperature of the flare, and $A_{flare}$ is the area covered by the flare. Since the ratio of the observed flare flux and stellar flux should be equal to the ratio of the flare luminosity and stellar luminosity, the following relationship can be used to estimate $A_{flare}$ as a function of time:
\begin{equation}
	\frac{\Delta F(t)}{F} = \frac{A_{flare}(t)\int R_{\lambda}B_{\lambda(T_{flare})}\mathrm{d}\lambda}{\pi R_*^2\int R_{\lambda}B_{\lambda(T_{eff})}\mathrm{d}\lambda},
\end{equation}
where $\frac{\Delta F}{F}$ is the amplitude of the flare, $\lambda$ is the wavelength, $R_{\lambda}$ is the \emph{Kepler} instrument response function, $B_{\lambda(T)}$ is the Planck function, and $R_*$ is the stellar radius. This assumes that the star can also be approximated by a blackbody radiator. To find the flare amplitude as a function of time, the flare flux, $\Delta F(t)$, was found first by masking out the flare data points and interpolating linearly over those points, then calculating the differences between the flare flux values and the interpolated flux values. The flare flux was then normalised by the mean flux in the vicinity of the flare, with the flare masked out. This method limits the impact of an underlying trend in the lightcurve on the calculated energy. The stellar temperature ($3748 \pm 112$\,K) and radius ($0.520^{+0.038}_{-0.061}$\,$R_{\odot}$) used to calculate the stellar luminosity were obtained by \citet{2014ApJS..213....5M} and \citet{2015ApJ...800...99T}, respectively. The flare emission was assumed to be a blackbody spectrum at $9000 \pm 500$\,K \citep{1992ApJS...78..565H, 2003ApJ...597..535H, Kretzschmar:2011cq}. The energies of the seven detected flares are given in Table \ref{tabk438flares}, and range from $(4 \pm 2) \times 10^{32}$ to $(1.4 \pm 0.6) \times 10^{33}$\,erg, where flares with energies greater than $10^{33}$\,erg are classified as superflares \citep{2012Natur.485..478M}. Solar flares typically do not reach beyond $10^{32}$\,erg. Since this star was observed for 4 years, on average a large flare occurs every \mytilde200 days. The strongest superflare seen is shown in Fig. \ref{figkep438flare}. 

\begin{table}
\begin{center}
\caption{Kepler-438 Flares}
\label{tabk438flares}
\begin{tabular}{lrr}
\hline
Quarter & Time 	&Energy \\
 & BJD - 2400000  &  $\times 10^{32}$\,erg \\
\hline
2 &55058.86			&$4 \pm 2$	\\
3 & 55141.14			&$7 \pm 3$	\\
6 & 55379.24			&$14 \pm 6$	\\
8 & 55572.92			&$8 \pm 4$	\\	
9 & 55701.39			&$11 \pm 5$	\\
10 & 55799.08			&$6 \pm 3$	\\
12 & 55974.30			&$4 \pm 2$	\\
\hline

\end{tabular}
\end{center}
\end{table}

We did not detect any significant flares in any of the other sample lightcurves. While converting this to limits on flare energies is non-trivial, we estimate upper limits on any flares present in each lightcurve by considering a flare with peak flux 5 standard deviations above the mean lightcurve level for each star. The standard deviations used to set the test flare amplitude were obtained from flattened versions of the PDC lightcurves. Flattening was performed by fitting a 3rd order polynomial to 2 day windows surrounding successive 0.2 day lightcurve segments. For each 0.2 day region, the fit is repeated 10 times, ignoring points more than $5\sigma$ discrepant from the previous fit. The polynomial is then removed from the 0.2 day region, and the process repeated for each 0.2 day region. These artificial flares were given durations of 7 long cadences, a typical flare duration, and given rapid rise times and an exponential decay. Seven cadences is significantly shorter than the 2 day window used for fitting flattening polynomials, and hence such flares would not be affected by the flattening procedure. The energy such flares would have as calculated in the same way as for Kepler-438, i.e. the upper limit on individual flare energy in each lightcurve, is given in Table \ref{tabflarelimits}.

\begin{table}
\begin{center}
\caption{Upper limits of flare energies where the flare would not be detected above the noise level. LC = Long Cadence, SC = Short Cadence.}
\label{tabflarelimits}
\begin{tabular}{llr}
\hline
Star	& Lightcurve & Energy Limit	\\
   &   & $\times 10^{32}$\,erg \\
\hline
Kepler-22	&LC	&6.5	\\
Kepler-22	&SC	&4.4	\\
Kepler-61	&LC	&3.1	\\
Kepler-61	&SC	&2.2	\\
Kepler-62	&LC	&4.5	\\
Kepler-62	&SC	&3.2	\\
Kepler-174	&LC	&5.4	\\
Kepler-174	&SC	&4.1	\\
Kepler-186	&LC	&1.4	\\
Kepler-186	&SC	&1.0	\\
Kepler-283	&LC	&7.6	\\
Kepler-296	&LC	&2.8	\\
Kepler-298	&LC	&7.9	\\
Kepler-298	&SC	&5.6	\\
Kepler-438	&LC	&1.6	\\
Kepler-440	&LC	&4.0	\\
Kepler-442	&LC	&5.6	\\
Kepler-443	&LC	&24.0	\\
KOI-4427	&LC	&6.9	\\
\hline

\end{tabular}
\end{center}
\end{table}

\subsection{Mass Loss Rates}
\label{sectmasslossrates}
Given that most of the stars considered here are cooler than the Sun, their habitable zones and hence the orbits of these planets are closer to their host stars than that of the Earth. As such it is worth considering the state of the stellar wind, and particularly the impact it may have on each planet (see next Section). To investigate the stellar mass loss rates we turn to the model of \citet{Cranmer:2011dm}, hereafter CS11. The CS11 model calculates stellar mass loss rates considering Alfv\'{e}n waves in the stellar atmosphere, with winds driven by hot coronal gas pressure and cool wave pressure in the extended chromosphere. The total mass loss rate is found by combining these two effects.

The CS11 model has the significant advantage of requiring only observable stellar parameters, namely the stellar radius, mass, luminosity, metallicity and rotation period. As many of these parameters are used simply to calculate parameters such as effective temperature and surface gravity, we find it preferable to recast the model into a form with inputs of effective temperature, log g, metallicity, stellar radius and rotation period. In this form all inputs are direct observables, except the stellar radius which is generally derived from spectra. We chose to use this model over the commonly used Parker wind model \citep{Parker:1958dn} due to the possibility of using these fundamental stellar parameters as inputs.

We use the input parameters given in Table \ref{tabinputparams} to populate the model. For stars where no good rotation period was found, we use a value of 30 days with an error of 1 day in order to estimate the mass loss rate. Although the other inputs such as temperature are well determined, we stress that in these cases the mass loss rates found are more of an estimate. In addition to mass loss rates, the CS11 model outputs Rossby numbers, Ro (the ratio of rotation period to the convective turnover time, $\tau_c$), which are given in Table \ref{tabderivedparams}. We do not give Rossby numbers for the stars with no good rotation period, as the Rossby number is a direct result of this period. The CS11 calculates $\tau_c$ and hence the Rossby number using the zero-age main sequence model of \citet{Gunn:1998iu}; see CS11 for a justification of this model over the various other methods of estimating $\tau_c$ \citep[e.g.][]{Landin:2010fi,Barnes:2010fc}. Errors are calculated through a Monte Carlo procedure whereby the model was run for 10000 iterations using normally distributed input variables. The mean and standard deviation of the output distributions give the values published here. We note that there can be significant discrepancies (up to \mytilde25\%) between different methods of estimating $\tau_c$, potential errors from which will feed into our Rossby numbers and hence values derived from them such as the gyrochronological ages. The errors given on these derived values do not incorporate these potential discrepancies in $\tau_c$.

\subsection{Stellar Ages}
We calculated gyrochronology ages using the formulation of \citet{Barnes:2010fc}. We assume a value of $1.1$\,d for the rotation period at time $t=0$ (following the calibrated Solar-mass model). Broad-band colour indices were taken from  the AAVSO Photometric All-Sky Survey \citep[APASS; ][]{Henden:2012wq}, accessed through the UCAC4 catalogue \citep{Zacharias:2013cf}. The convective turnover timescale $\tau_c$ was determined as described above. For each system we create Gaussian distributions for rotation period, $\tau_c$, and (B-V) colour, with mean and variance set to the known values and $1\sigma$ errors, respectively. These distributions were sampled $10^4$ times, and the gyrochronological age calculated for each sampling. Final ages for each system were taken to be the median of the appropriate set of results, with $1\sigma$ uncertainties set to the values which encompassed the central $68.3$\% of the data set.

\subsection{Planetary Magnetospheres}
The Earth's magnetosphere is a crucial element to the habitability of the planet. It shields the surface from energetic particles and radiation, and prevents excessive loss of atmosphere. Magnetospheric protection is important in assessing exoplanet habitability \citep[see e.g.][]{See:2014gk}. We consider the extent of the hypothetical magnetospheres of our sample planets, given an assumed field strength the same as the Earth's. In this work we are dealing with the radial distance of the magnetosphere from the planet's centre towards the star, the so-called magnetopause standoff distance. This distance is given where the planetary magnetospheric pressure balances the pressure from the star \citep{Griemeier:2004ez}, as in

\begin{equation}
r_{MP} = \left(\frac{\mu_0f_0^2M_E^2}{8\pi^2P_*}\right)^{\frac{1}{6}}
\label{eqnrmp}
\end{equation}
,where $r_{MP}$ represents the magnetopause standoff distance, $f_0$ is a form factor accounting for the non-spherical shape of the planet's magnetic field, here set to 1.16, $M_E$ is the Earth's magnetic moment, set to be $8\times10^{22}\textrm{Am}^2$, and $P_*$ represents the competing pressure from the star. The pressure on the planetary magnetosphere arises from stellar winds, as well as the magnetic and thermal pressures. We ignore the thermal pressure here, as the planets under consideration orbit relatively far from their host stars. The stellar wind pressure can be calculated from the mass loss rate as

\begin{equation}
P_{wind}(r) = \frac{\dot{M}v_{esc}}{4\pi r^2}
\end{equation}
,at orbital distance $r$ using the mass loss rate of the CS11 model $\dot{M}$, the stellar escape velocity $v_{esc}$ and following the prescription of \citet{See:2014gk}. At the orbital distances of our sample planets, the magnetic pressure can also be significant. This requires an estimate of the stellar magnetic fields. We use the empirically derived calibration between Rossby number and the large scale structure magnetic field strength of \citet{Vidotto:tt}. As all of our sample stars fall in the unsaturated regime with $Ro>0.1$ \citep[e.g.][]{James:2000gn,Pizzolato:2003ga,Wright:2011dj} the average field strength scales with $\left<B_{ZDI}\right> \propto Ro^{-1.38\pm0.14}$. We fix this relation using the saturated field strength found for early M stars in \citet{Vidotto:tt} of 50G at $Ro=0.1$. The average field strength is termed $B_{ZDI}$, as it is the component typically probed by the Zeeman Doppler Imaging technique.

The stellar magnetic field operates on both large and small scales; here the large scale field is the dominant contributor, as the small scale field falls off more rapidly with distance from the star \citep{Lang:2014hs}. The \citet{Vidotto:tt} calibration used gives the average large scale magnetic field strength at the stellar surface. A commonly used method to extrapolate stellar magnetic fields is the PFSS (potential field source surface) technique, which assumes that field strength falls off as $R^2$ beyond the source-surface radius $R_{SS}$ \citep[see e.g.][]{Vidotto:2013hz}. To calculate the average field strength $\left<B_{SS}\right>$ at $R_{SS}$, we assume for this first order estimate that the field follows a dipole structure between $R_*$ and $R_{SS}$, hence $\left<B_{SS}\right>$=$\left<B_{ZDI}\right>(R_*/R_{SS})^3$. We assume that $R_{SS}=2.5R_*$, as in \citet{Vidotto:2013hz}. Following this chain of relations, the magnetic pressure at orbital radius $r>R_{SS}$ is given by

\begin{equation}
P_{magnetic}(r) = \frac{\left<B_{SS}\right>^2}{2\mu_0} \left(\frac{R_{SS}}{r}\right)^4
\end{equation}
,with all variables measured in SI units. It should be noted that stellar magnetic fields typically contain higher order multipoles than the simple dipole term, which will increase the rate of dropoff between the stellar and source-surface radii. Without detailed observations we cannot constrain these stellar magnetic fields further, hence we use the simple dipole approximation. The resulting field strengths will then be overestimated, which will result in magnetospheric standoff radii smaller than the true values. This approach is conservative -- if the calculated magnetospheric standoff distances are large enough for the planets to maintain an Earth-like magnetosphere, then reduced magnetic pressure will only increase them.

In the cases where no good stellar rotation period (and hence Rossby number) is available, we assume a field $\left<B_{ZDI}\right>=5G$ to allow an estimate of the magnetic pressure. Using the above mass loss rates, Rossby numbers and scaling relations, we calculate planetary magnetospheric standoff radii through Equation \ref{eqnrmp}, with $P=P_{wind}+P_{magnetic}$. The resulting numbers are given in Table \ref{tabderivedparams}.

\subsection{X-ray Emission}
It is possible to estimate the X-ray emission of these stars from the obtained Rossby numbers, using the relation of \citet{Mamajek:2008jz}. All of our sample stars are in the unsaturated regime \citep[e.g.][]{Reiners:2009jb}, meaning that the X-ray luminosity of the star falls off with increasing Rossby number. The relation follows  

\begin{equation}
\log R_X = (-4.83\pm0.03)-(1.27\pm0.08)*(Ro-0.86)
\end{equation}
,where $R_X=L_X/L_{bol}$, the ratio of X-ray to bolometric luminosity, and $Ro$ is the Rossby number. The resulting values for log $R_X$ are given in Table \ref{tabderivedparams}. These stars all likely have X-ray luminosities too low to be easily observable with current technology, so in the absence of such data these estimates will prove useful in understanding the radiation environment of these planets.

\subsection{Results}  
All of the results from the above Sections are presented together in Table \ref{tabderivedparams} for clarity.

\begin{landscape}
\begin{table}
\begin{center}
\caption{Host Star Derived Parameters}
\label{tabderivedparams}
\begin{tabular}{lrrrrrrrrrrr}
\hline
Star & $P_{rot}$ (ACF) & $P_{rot}$ (GWS) &  $\left<S_{phot,5}\right>$ & Contrast & Age & $\dot{M}$ & Rossby & Log R$_X$ & L$_X$ &$B_{ZDI}$ & $R_{magneto}$ (planet)  \\ 
& d & d &  ppm&  & Gyr & $M_\odot yr^{-1}$& & & L$_{X,\odot}$$^{e}$ & G & $R_\oplus$   \\
\hline
Kepler-22 & - & - &  312&2.47& - & $(2.5\pm0.6)x10^{-14}$$^{a}$& - &  - & - & 5$^{b}$ & 8.7 (b)\\[2pt]
Kepler-61 &$35.55 \pm 0.38$ & $34.8 \pm 5.0$ &  1680 &1.45& $7.5\pm0.5$ & $(2.1\pm1.2)x10^{-16}$ & $0.91\pm0.07$  &  $-4.89\pm0.09$ & $4.0\pm1.3$ & 2.4 & 12.7(b)\\[2pt]
Kepler-62 & $39.77\pm0.44$ & $37.3\pm13.3$ & 405& 1.11& $14.6 \pm 0.6$ &$(1.4\pm0.2)x10^{-15}$& $1.61\pm0.06$  & $-5.78\pm0.12$ & $1.2\pm0.5$ & 1.1 & 11.1 (e), 13.2 (f)\\[2pt] 
Kepler-174 & - & - &  445 & 1.47 & - &$(2.6\pm1.4)x10^{-15}$$^{a}$& - & - & - & 5$^{b}$ & 11.6(d) \\[2pt]
Kepler-186 &$34.27\pm0.07$ & $33.8\pm 4.2$ &  3270 & 2.32 & $6.0 \pm 0.2$& $(4.6\pm1.5)x10^{-17}$ & $0.75\pm0.03$ & $-4.69\pm0.06$ & $3.4\pm0.8$ & 3.1& 19.2(f)\\[2pt]
Kepler-283 &$18.27 \pm 0.02$ & $18.1 \pm 2.2$ & 2620 & 1.52 &$2.4\pm0.1$ & $(1.7\pm1.0)x10^{-15}$ &$0.54\pm0.03$&$-4.42\pm0.05$ & $13.4\pm2.6$ & 4.9 & 9.9(c)\\[2pt]
Kepler-296&$36.11\pm 0.13$& $35.0\pm4.8$ &  2800 & 1.36 & $6.6\pm0.4$ & $(2.9\pm2.3)x10^{-17}$ & $0.78\pm0.05$&$-4.73\pm0.08$ & $2.6\pm1.1$ & 2.9 &13.7(e), 16.7(f)\\[2pt]
Kepler-298 & - & - &  967 & 1.51 & - & $(8.2\pm4.4)x10^{-16}$$^{a}$& - & - & - & 5$^{b}$ & 10.7(d)\\[2pt]
Kepler-438 &$37.04 \pm 0.08$& $36.8\pm4.5$ &  3260 & 1.90 & $7.1\pm0.4$  & $(3.2\pm1.8)x10^{-17}$ &$0.81\pm0.05$& $-4.77\pm0.08$& $2.8\pm1.0$ & 2.8 & 13.2(b)\\[2pt]
Kepler-440 &$17.61\pm 0.02$& $17.3\pm2.1$ &  2420 & 1.37 & $2.0\pm0.1$ & $(9.8\pm5.8)x10^{-16}$ &$0.47\pm0.04$&$-4.33\pm0.06$ & $16.9\pm4.4$ & 5.9 & 9.6(b)\\[2pt]
Kepler-442 & - & - &  736 & 1.50 & - & $(7.7\pm2.3)x10^{-16}$$^{a}$ & - & - & - & 5$^{b}$ & 12.0(b)\\[2pt]
Kepler-443 & - & - &  1250 & 1.25 & - & $(2.4\pm0.6)x10^{-15}$$^{a}$& - & - & - & 5$^{b}$ & 10.6(b)\\[2pt]
KOI-4427 & - & - &  1410 &1.44 & - &$(6.3\pm3.6)x10^{-17}$$^{a}$ & - & - & - & 5$^{b}$ & 17.7(b)\\[2pt]
\hline
Solar$^{c}$ & 25.0 & - & 166.1 & 2.7  & 4.569 & $2x10^{-14}$ &  1.577  &  -6.24 & 1 & 1.89--3.81 & 9.50(Earth)$^{d}$ \\
\hline 
\multicolumn{12}{l}{$^a$ Using rotation period of 30 days and error of 1 day}\\
\multicolumn{12}{l}{$^b$ Example value used as no rotation period available}\\
\multicolumn{12}{l}{$^c$ Solar values taken from \citet{Mathur:2014cz,Vidotto:tt,Cohen:2011ht,Bouvier:2010iv} and references therein.}\\
\multicolumn{12}{l}{$^d$ Planetary magnetosphere extent for the Earth when calculated using the same method as the sample, for reference. The true standoff radius for the Earth varies} \\
\multicolumn{12}{l}{between 9.7--10.3 $R_\oplus$. }\\
\multicolumn{12}{l}{$^e$ Using Log L$_{X,\odot}=20.05$W. L$_X$ values calculated from Log R$_X$ and values from Table \ref{tabinputparams}.}\\
\end{tabular}
\end{center}
\end{table}
\end{landscape}

\section{Discussion}
\subsection{Rotation Periods}
\label{discprot}
We find clear rotation signals for 6 stars, plus a present but less clear signal for Kepler-62. Each of these periods has been previously published in either the discovery papers or those cited below, but often in a number of different locations, using different methods and sets of data. Here we present what we believe to be the most comprehensive study of the rotation of these stars to date, using the entire \emph{Kepler} dataset, alternative detrending methods, and a combination ACF and wavelet analysis. 

Previous studies of rotation in \emph{Kepler} stars include \citet{McQuillan:2013jw}, who used the ACF technique, data from quarters 3-14 and give consistent periods to ours for Kepler-61, Kepler-186, Kepler-283 and Kepler-296. Rotation periods for these stars were also given by \citet{Walkowicz:wh} and \citet{Reinhold:2013iz} using Lomb-Scargle periodogram techniques \citep{Lomb:1976bo,Scargle:1982eu}, although both of these studies give half the true rotation period for Kepler-61. Moreover, in \citet{Reinhold:2013iz} while the periods given for the other 3 stars are close to ours, they are not consistent within the stated errors. \citet{Reinhold:2013iz} also give periods for Kepler-440 and KOI-4427. For KOI-4427, we find that their period (26.6 days) matches some active regions on our wavelet plot, but by no means represents a dominant or persistent signal. As such we cannot claim agreement for KOI-4427. None of our sample appear in the \citet{Nielsen:2013cp} study of \emph{Kepler} rotation periods.

It is interesting that all of our detected rotation periods are relatively long for \emph{Kepler} stars, near the upper end of the range of periods found in \citet{McQuillan:2014gp}. This may be a result of observation bias, as longer rotation period stars tend to be less active and so in principle easier to detect transits within. However, as the next section shows, our sample stars are all quite active. The stars for which we do not detect rotation periods may be older, inactive stars, limiting the effect of any spot modulation which would allow us to detect their rotation signal. As such in terms of habitability, they may be preferred targets. In general these stars have relatively low $\left<S_{phot,5}\right>$ values. In the more active cases, photometric variability can arise from sources other that starspots, such as in shorter timescale `flicker' .

\subsection{Stellar Photometric Activity}
\label{discactivity}
The solar value of photometric activity when measured using $\left<S_{phot,5}\right>$ is 166.1ppm, with a maximum value of 285.5ppm \citep{Mathur:2014cz}. Every star in our sample has a mean photometric activity level above the solar maximum, in some cases by an order of magnitude. They are also significantly more active than stars in similar studies, including \citet{Mathur:2014cz} and \citet{Garcia:2014ds}. This is to be expected, as our sample is dominated by K stars as opposed to the hotter targets studied previously (including the Sun). Cool stars are typically more active, but also represent easier targets for habitable planet detection due to their closer habitable zones. We note that the S$_{phot,k}$ index does not consider stellar inclination, and as such if these stars were inclined to the line of sight their underlying photometric activity could be larger.

We do not see any strong evidence for magnetic cycles in our sample stars, which is unsurprising given the 4 year data baseline. This has the consequence that the activity indexes we measure are in effect `snapshots' of the overall value, captured at a certain phase of the potential activity cycle. It is quite possible that activity will increase (or decrease) at other phases. Disentangling this effect would require more data spanning several years. Although it depends on target selection, there is the possibility of the TESS \citep{Ricker:2014fy} or PLATO \citep{Rauer:2014kx} satellites revisiting these targets at a later time.

It is unclear what direct effect activity can have on habitability, beyond the well-known difficulties it introduces in searching for small planets \citep[e.g.][]{Cegla:2012bq,Haywood:2014hs}. The connection between photometric and spectroscopic activity is not clear, beyond a small sample studied by \citet{Bastien:2013dt}. It seems likely that increased photometric activity will lead to increased spectroscopic activity, and hence difficulty in securing radial velocities of low mass rocky planets. Activity tends to be a marker of increased CMEs and flares, which can play a role in habitability in potentially stripping planetary atmospheres \citep[e.g.][]{Lammer:2007kh}. We investigate some of these stellar properties more directly in this work, but flares occurring at rates less than the 4 year \emph{Kepler} data would not be seen. If energetic enough they could still pose some issue for their resident planets. Strongly active, flaring stars such as Kepler-438 can destroy atmospheric biomarkers in a planet's atmosphere \citep{JohnLeeGrenfell:2012fa,Grenfell:2014gy}, with implications for future searches for life.

\subsection{Flares}
The detection of superflares in the lightcurve of Kepler-438 could pose some interesting issues for habitability. An investigation into the impact of a powerful flare from an M dwarf on an Earth-like exoplanet in the habitable zone was made by \citet{2010AsBio..10..751S}. They found that the increase in UV and X-ray emission associated with a flare would not have a significant impact on habitability, since X-rays could not penetrate beyond the upper atmosphere, and UV radiation at the surface would only reach slightly higher levels than on Earth. Any temperature increase due to ozone depletion would also be minor. In the absence of a planetary magnetic field, however, an increase in the flux of energetic charged particles associated with large flares could potentially be damaging to life. 

On the Sun another phenomenon associated with flares is a coronal mass ejection (CME), where a large amount of coronal material is expelled, often at high speeds \citep{Gosling:1976cw}. The likelihood of a CME occuring increases with more powerful flares \citep{Kahler:1992kj,Yashiro:2005bx}, however the relationship between flares and CMEs is complex, with some CMEs occuring without an associated flare \citep{Munro:1979bm}. It is possible that CMEs occur on other stars that produce very energetic flares, which could have serious consequences for any close-in exoplanets without a magnetic field to deflect the influx of energetic charged particles. Since the habitable zone for M dwarfs is relatively close in to the star, any exoplanets could be expected to be partially or completely tidally locked. This would limit the intrinsic magnetic moments of the planet, meaning that any magnetosphere would likely be small. \citet{Khodachenko:2007hm} found that for an M dwarf, the stellar wind combined with CMEs could push the magnetosphere of an Earth-like exoplanet in the habitable zone within its atmosphere, resulting in erosion of the atmosphere. Following on from this, \citet{Lammer:2007kh} concluded that habitable exoplanets orbiting active M dwarfs would need to be larger and more massive than Earth, so that the planet could generate a stronger magnetic field and the increased gravitational pull would help prevent atmospheric loss. 

Given Kepler-438b's relatively close orbit to its M dwarf host, the eventual effect of these flares is strongly dependent on the size and strength of its magnetic field. The magnetospheric extent calculated in Table \ref{tabderivedparams} assumes an Earth-like magnetic moment, which may not be achievable for such a planet. If it has a higher than expected mass this would help prevent atmospheric mass loss, but we cannot determine this with current observations. In any case, the high rate of strong flares (\mytilde every 200 days) must be taken into account when considering Kepler-438b as an Earth-like planet.

In all other cases except Kepler-443, our calculations of upper limits on the flare energy rule out flares with greater than $10^{33}$ ergs, the threshold for flares being classified as `superflares'. Hence in these cases the situation may be much improved relative to Kepler-438. We reiterate however that the most significant factor determining the effect such flares and connected CMEs may have is the strength of the planetary magnetospheres, something almost completely unknown. 

\subsection{Planetary Magnetospheres}
Encouragingly all of our sample planets are able to sustain magnetospheres of near Earth size or greater, assuming Earth strength magnetic moments. Even the smallest magnetopause standoff distance in our sample (Kepler-440b, with $R_\textrm{magneto}=9.6R_\oplus$, ignoring stars without well-defined Rossby numbers) is well above the Paleoarchaen Earth magnetopause standoff distance of \mytilde $5R_\oplus$ 3.4 Gyr ago \citep{Tarduno:2010gq}. This implies that subject to the generation of a planetary magnetic field similar to the Earth's in strength, magnetic protection should be effective for these planets at the present time. However, it is worth considering that magnetic activity and stellar winds are both expected to decline with age \citep[e.g.][]{Reiners:2012ce}. As such earlier in the system's histories the extent of the magnetic field may not have been so large, and atmospheric stripping could become a possibility. However, \citet{See:2014gk} find that for stars slightly more massive than M dwarfs (i.e. the K dwarfs which comprise the bulk of this sample) planetary magnetic protection is more effective over the stellar lifetime. Our results support this conclusion, albeit at only one epoch. In terms of the generation of planetary magnetic fields, \citet{Driscoll:2015ux} studied tidal dissipation in potentially habitable exoplanets, and found that such dissipation can have significant effects on a planet's core. If the core solidifies, the planetary dynamo will become ineffective, making the generation of a strong planetary magnetic field impossible. Core-mantle interactions can also weaken the generated field.

\subsection{Stellar Mass Loss Rates}
Most of the mass loss rates reported in Table \ref{tabderivedparams} are an order of magnitude or more lower than the solar value. Calculating the solar mass loss rate using the CS11 model gives a value of $3.6x10^{-14} M_\odot yr^{-1}$, a slightly overestimated but reasonable result which supports the validity of the model. The lower values found for our sample stars are a result of their longer rotation periods as compared to the Sun. The derived mass loss rates here are again often orders of magnitude lower than measured mass loss rates for other stars (see the list of such measurements in CS11, and references therein). We note the small number of measured rates however, which as a sample have shorter rotation periods and hence larger Rossby numbers. If the true mass loss rates of our sample stars are in fact higher, it will decrease the calculated planetary magnetopause standoff distances.

\subsection{X-ray Emission}
All of our calibrated values for X-ray emission are higher than the solar value of $\log R_X = -6.24$ \citep{Judge:2003gp}, generally by orders of magnitude. This is to be expected for these later type stars. Given that the habitable zone of such cooler stars lies closer to the star, the reduced planetary orbital distance and increased X-ray emission will lead to a significantly higher incident flux of X-rays at the planet's surface, though in the presence of an atmosphere much of this would be attenuated before reaching the surface \citep{Cnossen:2007bh}. Higher stellar X-ray flux is also associated with increased CME and planetary atmospheric stripping \citep{Lammer:2007kh}, seriously affecting habitability. We expect these values to be useful in future modelling of the \emph{Kepler} habitable exoplanet atmospheres.

\section{Conclusion}
We have presented a study of the lightcurves of the host stars of the most Earth-like \emph{Kepler} planets known to date. We also present the detection of superflares on Kepler-438, the most Earth-like exoplanet known to date according to the Earth Similarity Index, with energetic flares occurring every \mytilde200 days. These may have strong consequences for planetary habitability, depending on the strength of the planet's magnetosphere. We have also presented the most comprehensive study of stellar rotation on these stars to date, using the whole \emph{Kepler} dataset as well as an alternative detrending method. Using these rotation periods and previously published stellar parameters, we have calculated the photometric activity indices, mass loss rates, gyrochronological ages, Rossby numbers, X-ray fluxes and magnetic pressure exerted by these stars. This leads to an estimate of the magnetospheric standoff distance for the planets, assuming Earth-like magnetospheres, showing that in principle these planets could all maintain an Earth-sized magnetosphere. We stress that the parameters not calculated directly from the lightcurves (i.e. mass loss rates, ages, Rossby numbers, Xray fluxes and magnetic field values) are model based estimates, and should be treated as such. All of the directly and indirectly calculated parameters may come to bear in assessing planetary habitability, and in future detailed studies of these systems. 

\section*{Acknowledgments}

The authors would like to thank V See for very helpful comments made on the manuscript, and E Guinan for a thorough and helpful review. We would also like to thank the \emph{Kepler} team for their hard work, without which the data which makes this research possible would not exist. Funding for the Stellar Astrophysics Centre is provided by The Danish National Research Foundation (Grant agreement no.: DNRF106). The research is supported by the ASTERISK project (ASTERoseismic Investigations with SONG and \emph{Kepler}) funded by the European Research Council (Grant agreement no.: 267864).

\bibliography{papers_161015}
\bibliographystyle{mn2e_fix}

\appendix

\section{Wavelet Plots}
Here we show the wavelet plots associated with the lightcurves used to produce the periods given in Table \ref{tabderivedparams}. For stars where no conclusive period was given, the wavelet plot of the 100 day KASOC lightcurve is shown.

\begin{figure*}[h]
\resizebox{\hsize}{!}{\includegraphics{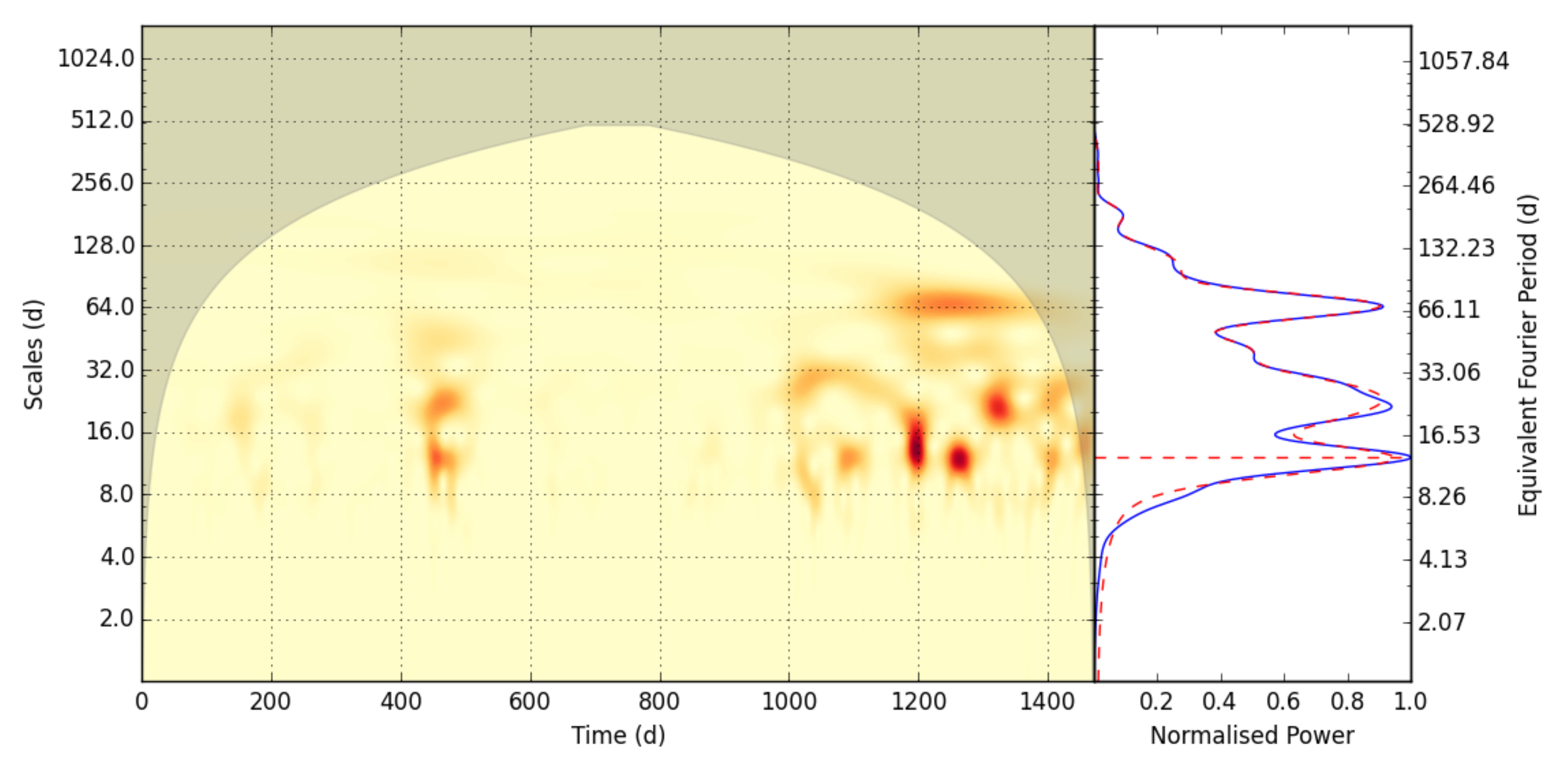}}
\caption{Contour wavelet plot for Kepler-22, calculated using the KASOC 100d lightcurve. The global wavelet spectrum is plotted to the right and fit by a sum of Gaussians (dashed line). The horizontal dashed line in the GWS represents the maximum peak seen.}
\label{figkep22wav}
\end{figure*}

\begin{figure*}
\resizebox{\hsize}{!}{\includegraphics{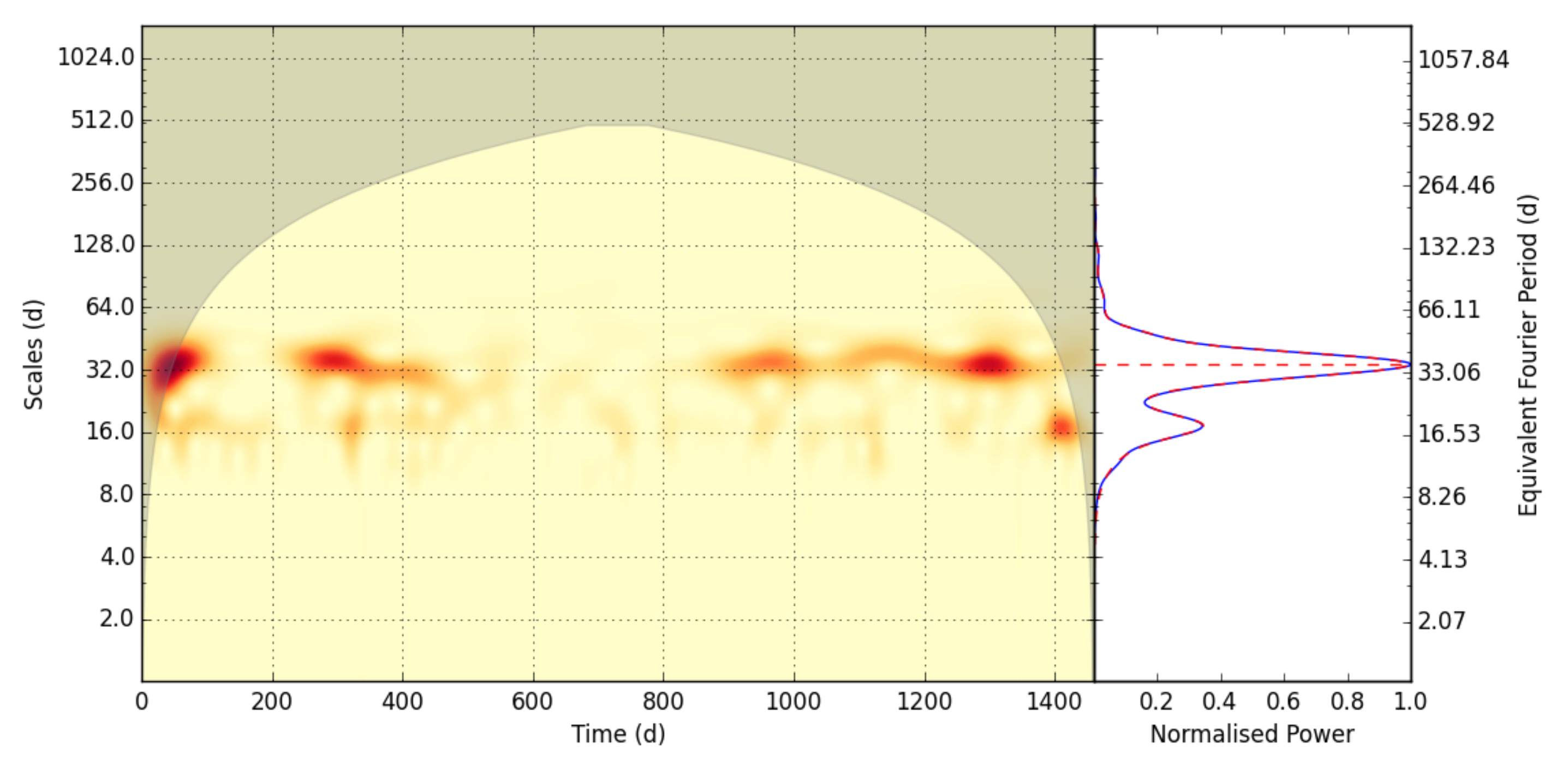}}
\caption{Contour wavelet plot for Kepler-61, calculated using the KASOC 50d lightcurve. See Fig. \ref{figkep22wav} for description.}
\end{figure*}

\begin{figure*}
\resizebox{\hsize}{!}{\includegraphics{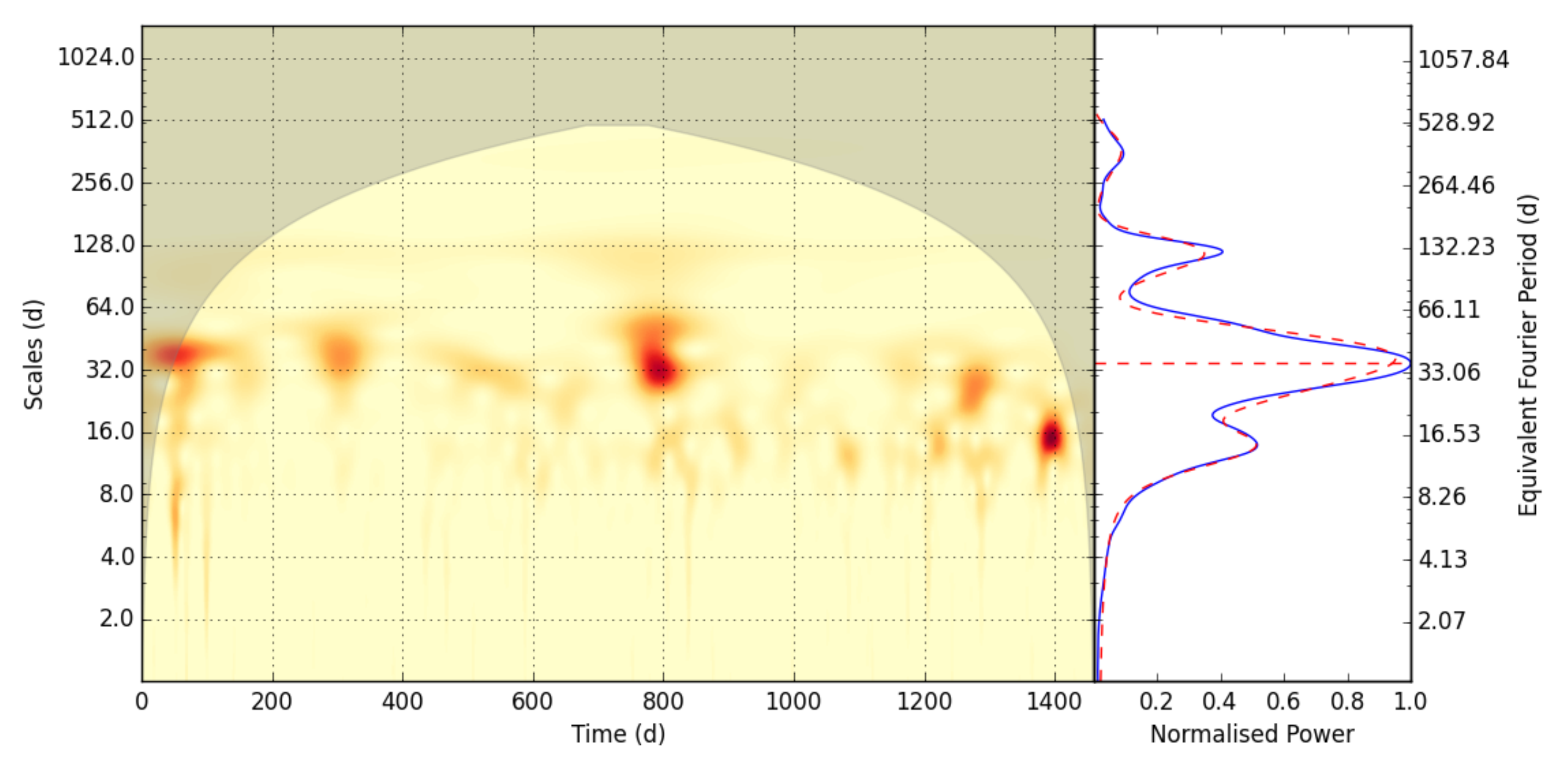}}
\caption{Contour wavelet plot for Kepler-62, calculated using the KASOC 50d lightcurve. See Fig. \ref{figkep22wav} for description.}
\end{figure*}

\begin{figure*}
\resizebox{\hsize}{!}{\includegraphics{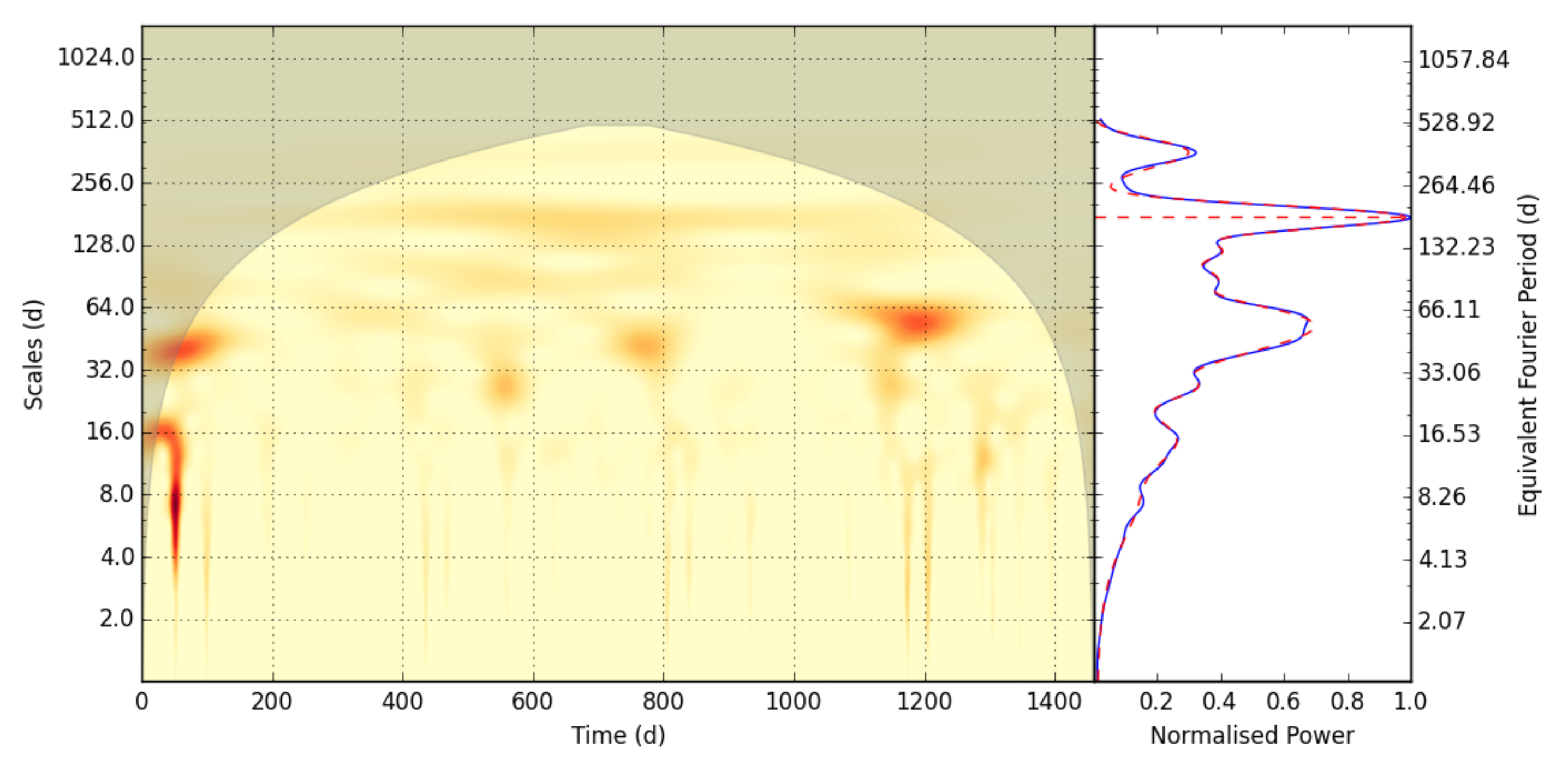}}
\caption{Contour wavelet plot for Kepler-174, calculated using the KASOC 100d lightcurve. See Fig. \ref{figkep22wav} for description. For Kepler-174 data points near gaps of greater than 0.8 days were cut due to excessive gap related discontinuities.}
\end{figure*}

\begin{figure*}
\resizebox{\hsize}{!}{\includegraphics{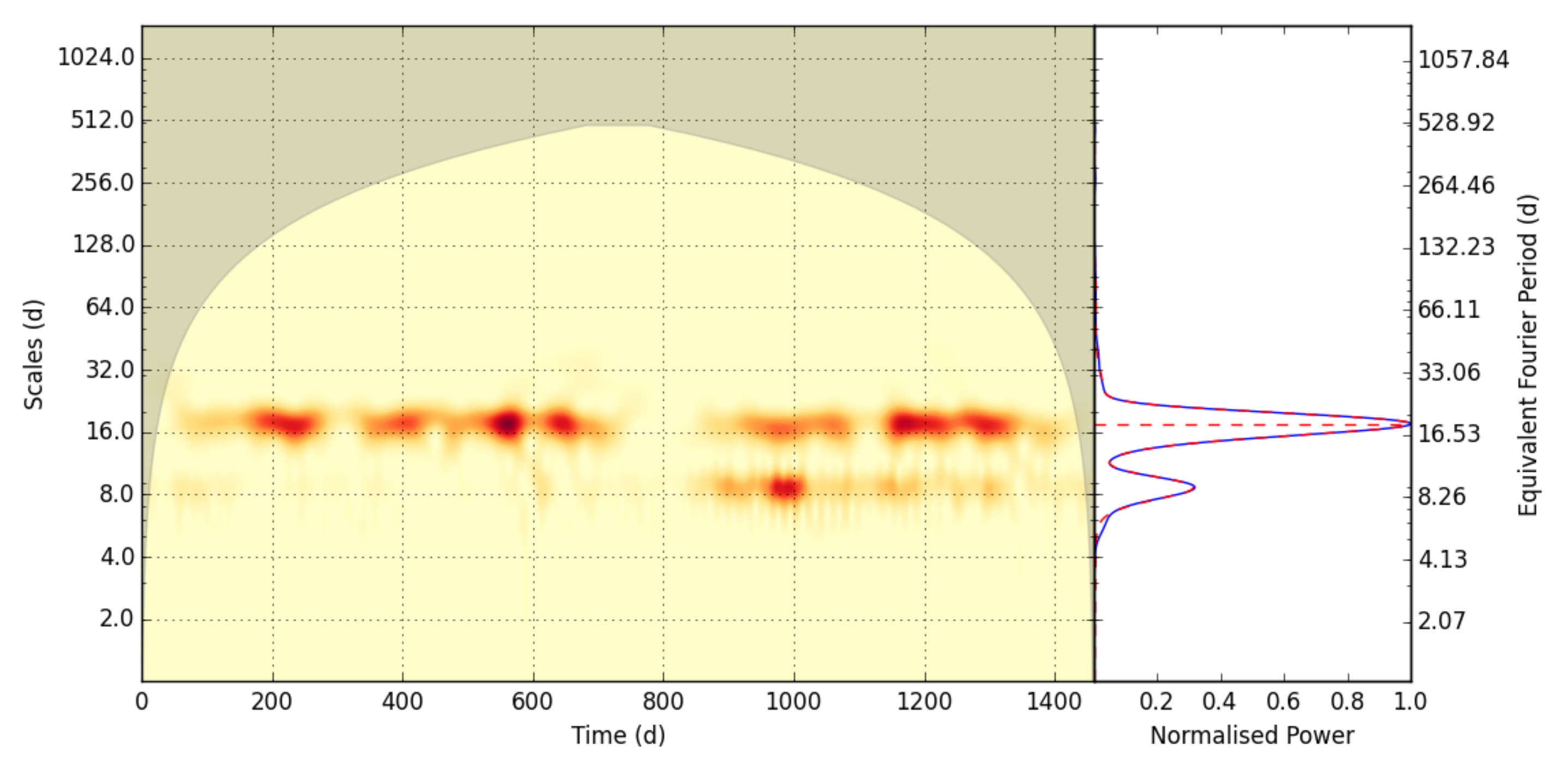}}
\caption{Contour wavelet plot for Kepler-283, calculated using the KASOC 30d lightcurve. See Fig. \ref{figkep22wav} for description.}
\end{figure*}

\begin{figure*}
\resizebox{\hsize}{!}{\includegraphics{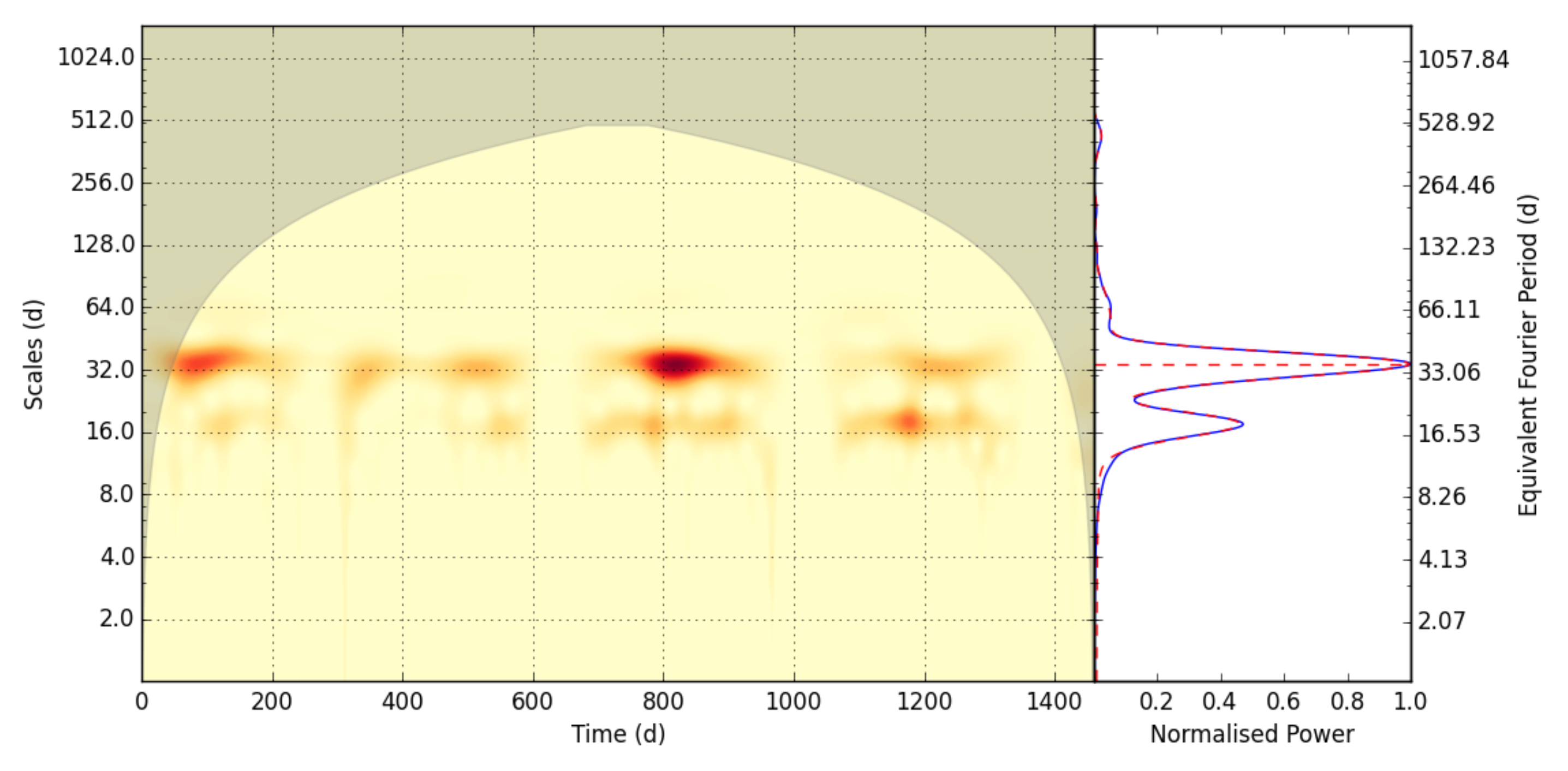}}
\caption{Contour wavelet plot for Kepler-296, calculated using the KASOC 30d lightcurve. See Fig. \ref{figkep22wav} for description.}
\end{figure*}

\begin{figure*}
\resizebox{\hsize}{!}{\includegraphics{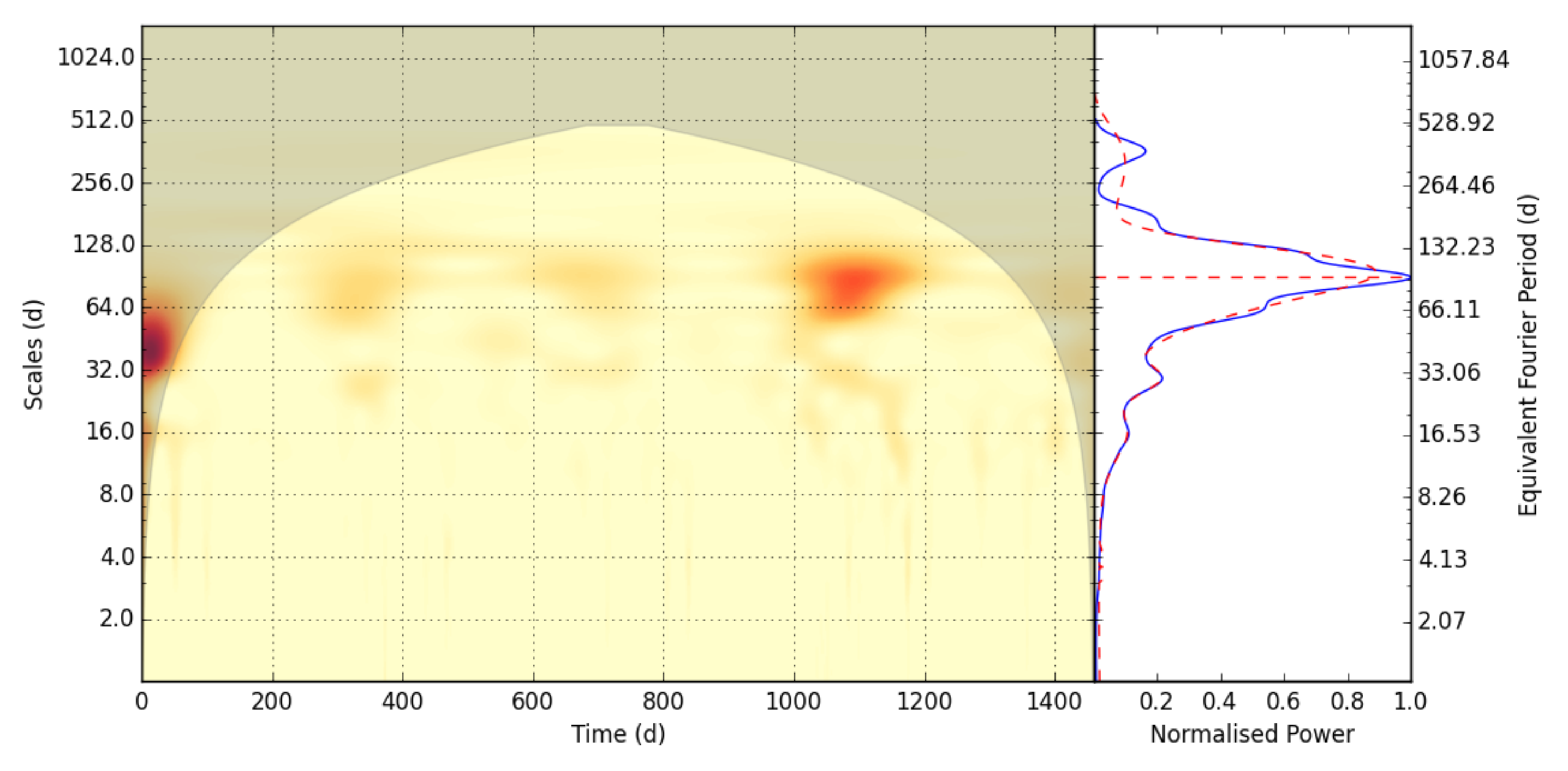}}
\caption{Contour wavelet plot for Kepler-298, calculated using the KASOC 100d lightcurve. See Fig. \ref{figkep22wav} for description.}
\end{figure*}

\begin{figure*}
\resizebox{\hsize}{!}{\includegraphics{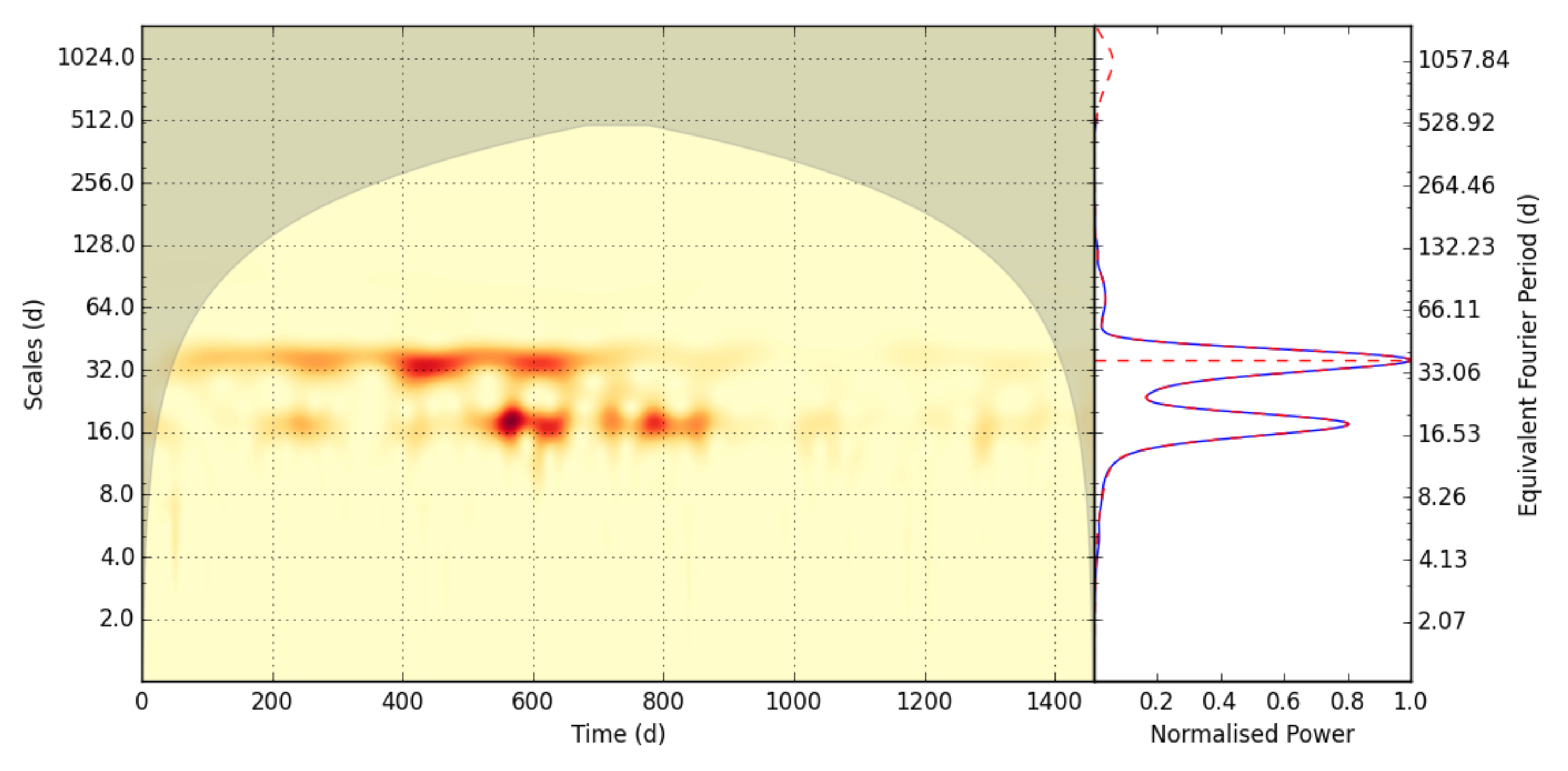}}
\caption{Contour wavelet plot for Kepler-438, calculated using the KASOC 30d lightcurve. See Fig. \ref{figkep22wav} for description.}
\end{figure*}

\begin{figure*}
\resizebox{\hsize}{!}{\includegraphics{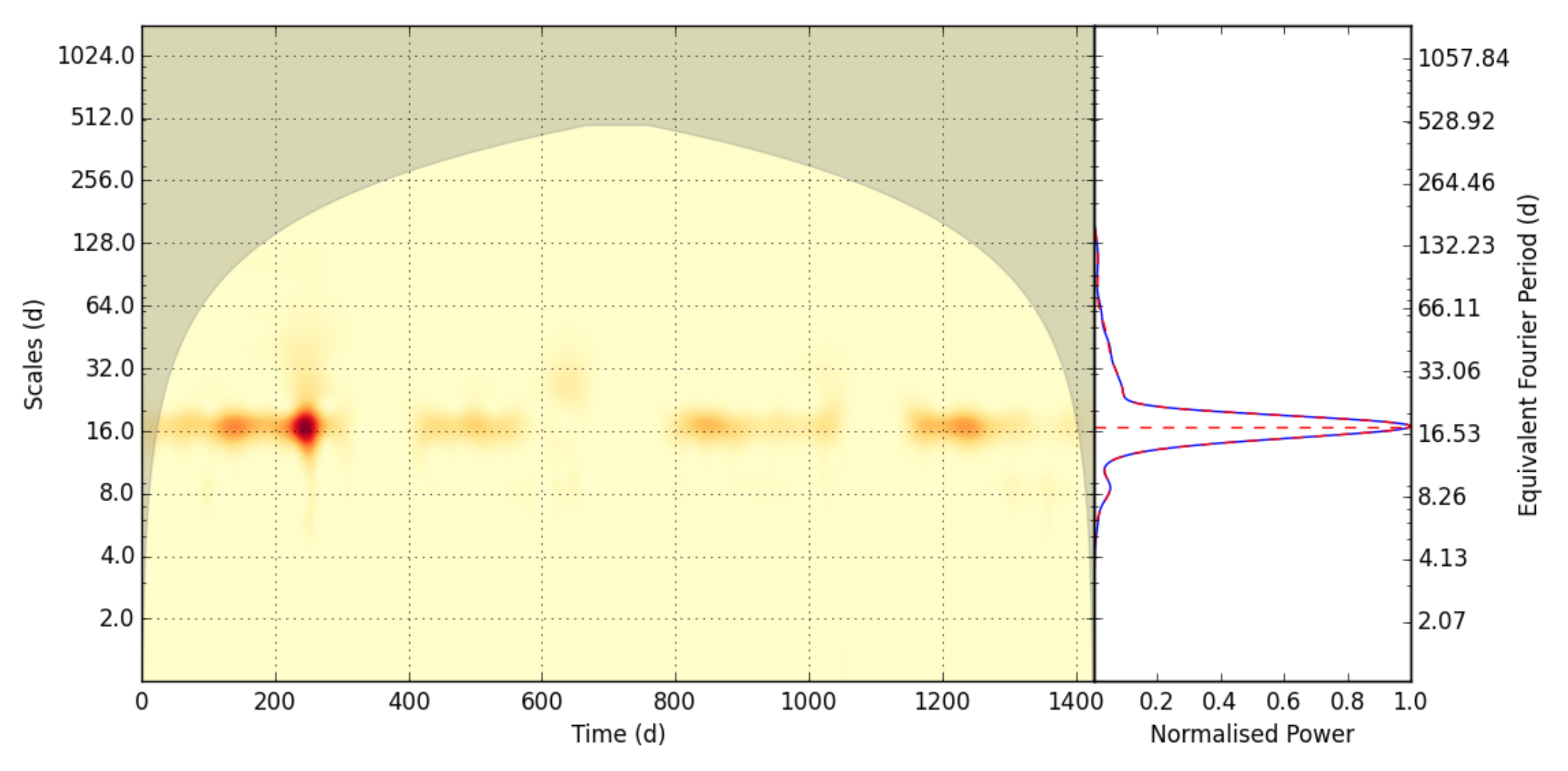}}
\caption{Contour wavelet plot for Kepler-440, calculated using the KASOC 30d lightcurve. See Fig. \ref{figkep22wav} for description.}
\end{figure*}

\begin{figure*}
\resizebox{\hsize}{!}{\includegraphics{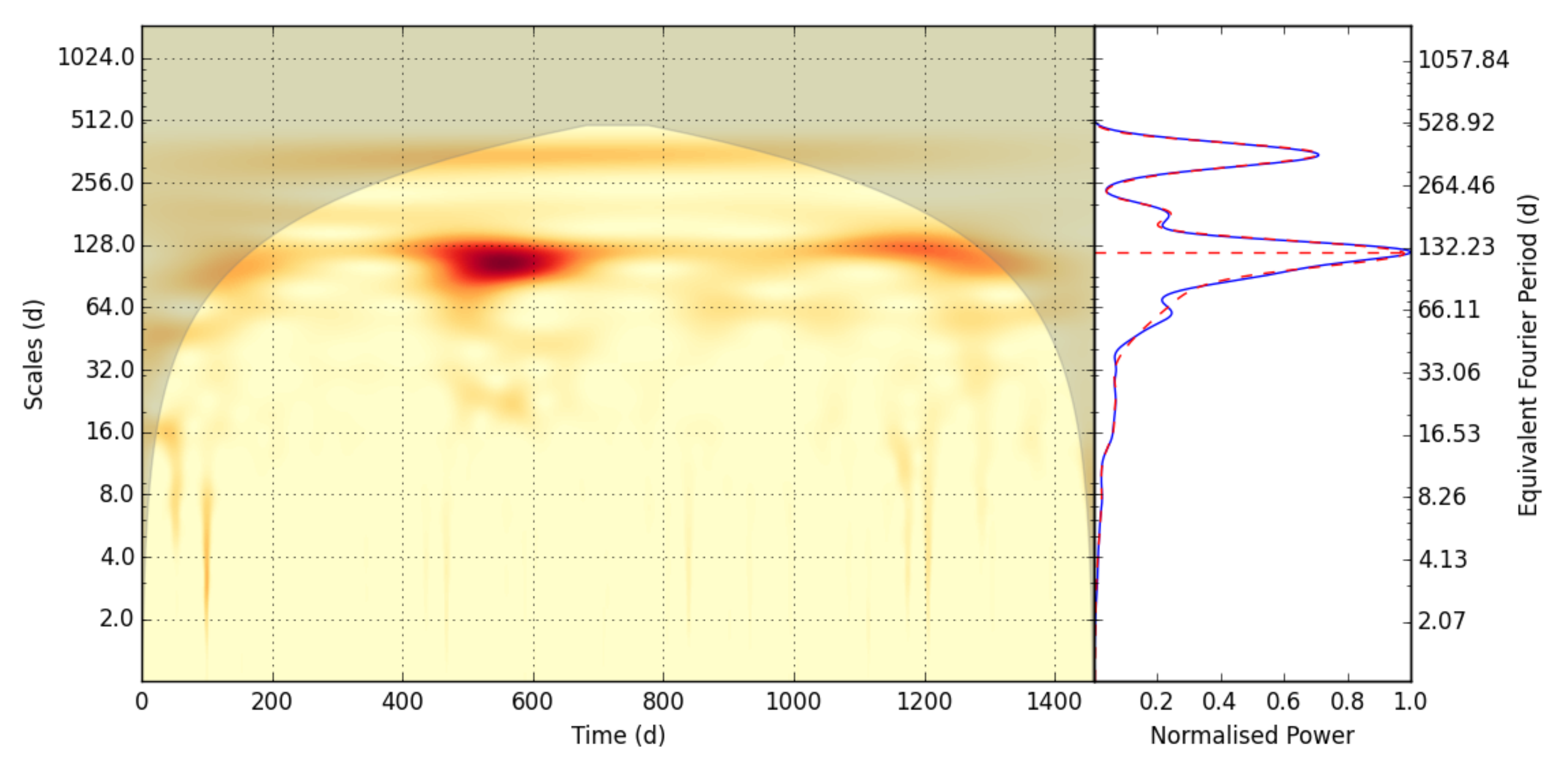}}
\caption{Contour wavelet plot for Kepler-442, calculated using the KASOC 100d lightcurve. See Fig. \ref{figkep22wav} for description.}
\end{figure*}

\begin{figure*}
\resizebox{\hsize}{!}{\includegraphics{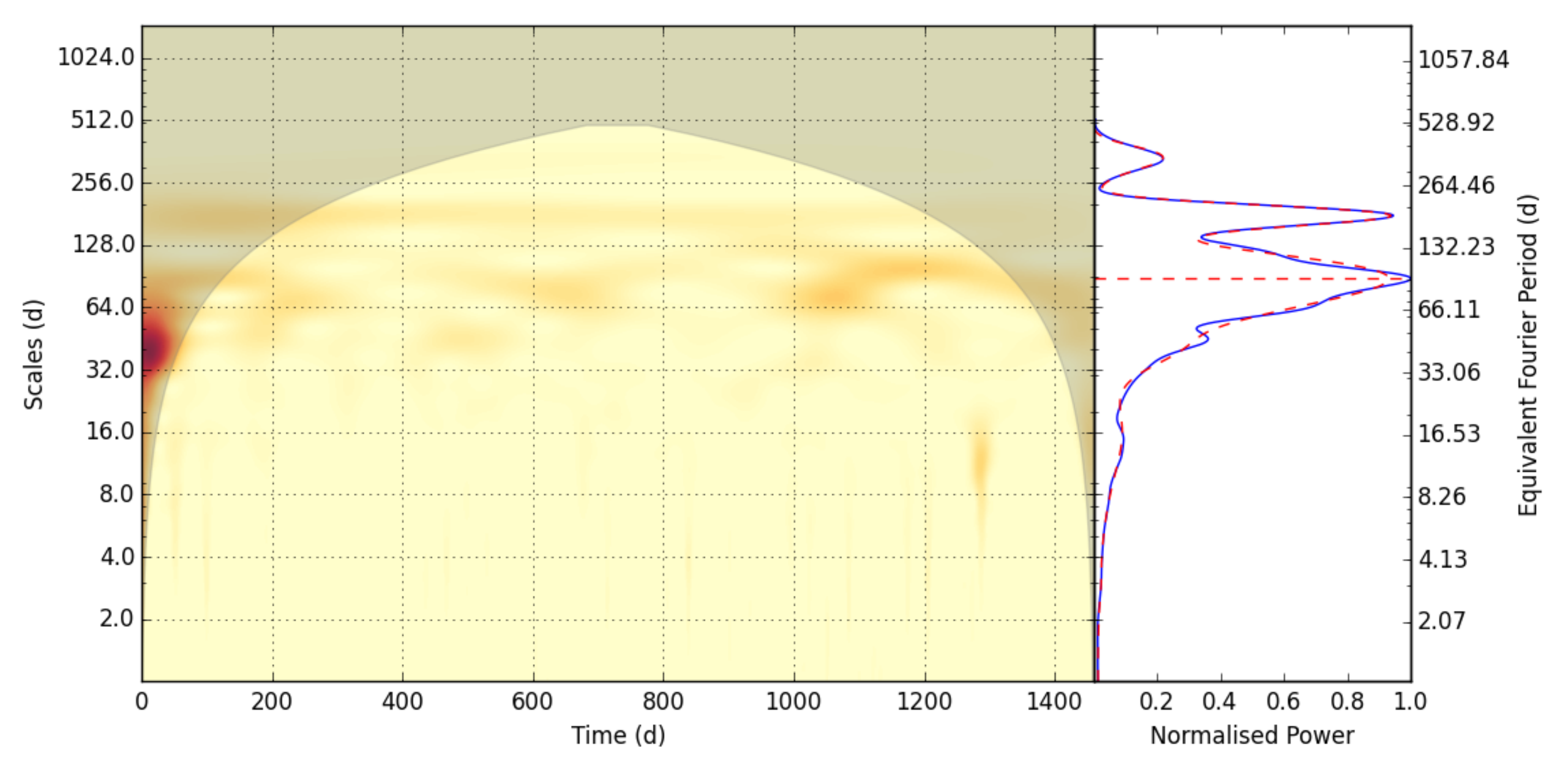}}
\caption{Contour wavelet plot for Kepler-443, calculated using the KASOC 100d lightcurve. See Fig. \ref{figkep22wav} for description.}
\end{figure*}

\begin{figure*}
\resizebox{\hsize}{!}{\includegraphics{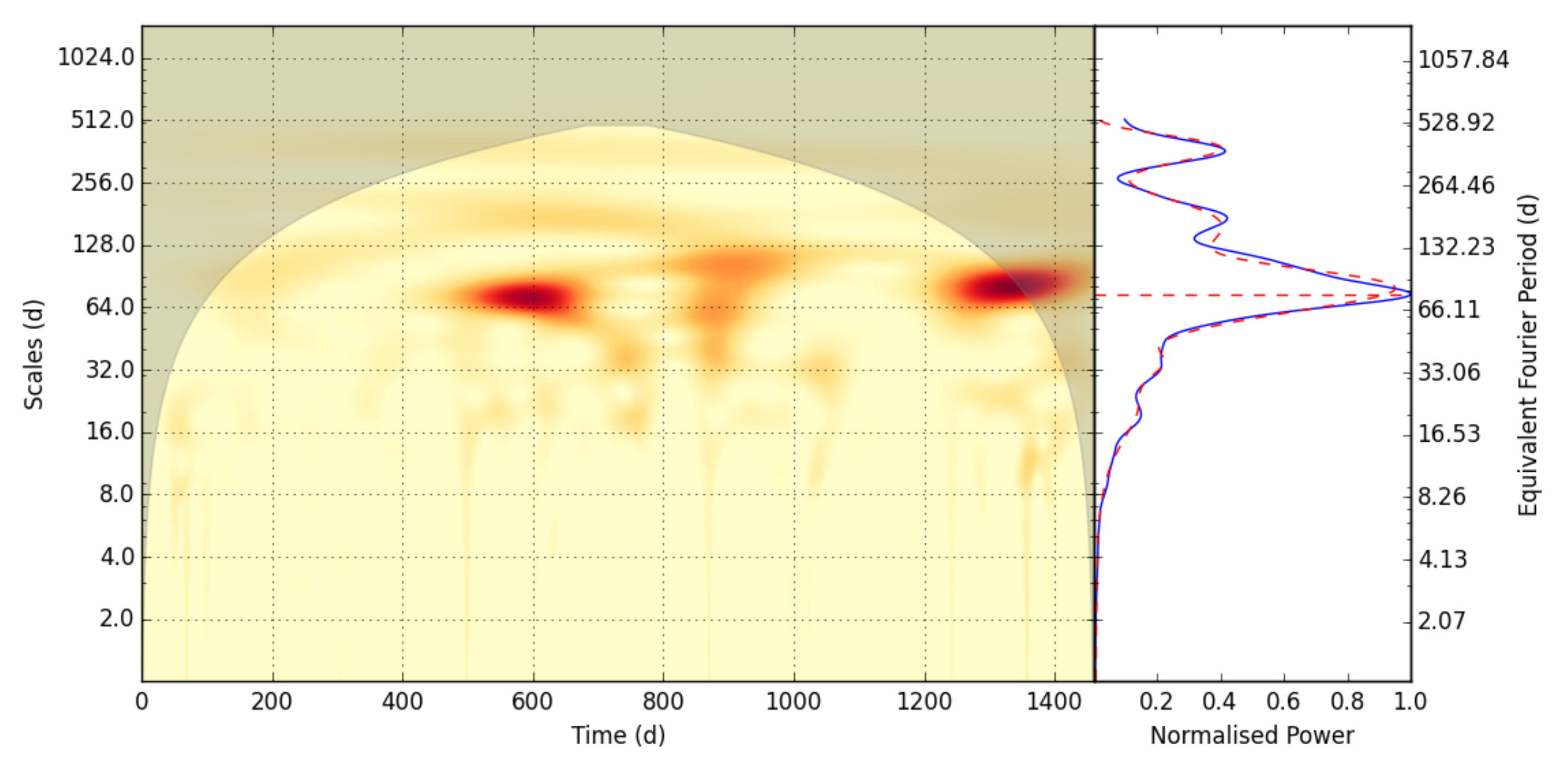}}
\caption{Contour wavelet plot for KOI-4427, calculated using the KASOC 100d lightcurve. See Fig. \ref{figkep22wav} for description.}
\end{figure*}


\end{document}